\documentclass[11pt]{article} 
\pdfoutput=1
\usepackage{amsmath,amssymb,amsfonts,cite,fullpage,graphicx,multirow,nicefrac,slashed}
\usepackage{color}
\definecolor{colorLink}{rgb}{0.9,0,0} % red
\definecolor{colorCite}{rgb}{0,0.7,0} % green
\definecolor{colorURL} {rgb}{0,0,0.8} % navy
\usepackage[colorlinks=true,linktocpage=true,linkcolor=colorLink,citecolor=colorCite,urlcolor=colorURL]{hyperref}

\newcommand{\be}{\begin{equation}}
\newcommand{\ee}{\end{equation}}
\newcommand{\bea}{\begin{eqnarray}}
\newcommand{\eea}{\end{eqnarray}}

\newcommand{\nn}{\nonumber}

\title{
\begin{flushright}
\vspace{-1cm}
\normalsize CERN-TH-2018-032
\end{flushright}
\vspace{1cm}
\LARGE\bf Exponentially Light Dark Matter from Coannihilation}
\author{\large Raffaele Tito D'Agnolo$^a$, Cristina Mondino$^b$, Joshua T. Ruderman$^{b,c}$, and Po-Jen Wang$^b$}
\date{\it \small $^a$SLAC National Accelerator Laboratory, 2575 Sand Hill Road, Menlo Park, CA, 94025, USA.\\ 
  \vspace{0.8em} $^b$Center for Cosmology and Particle Physics, Department of Physics, \\ New York University, New York, NY 10003, USA. \\
     \vspace{0.8em} $^c$Theoretical Physics Department, CERN, Geneva, Switzerland.
  }

\begin{document}

\begin{titlepage}

\maketitle

\begin{abstract}
\noindent
Dark matter may be a thermal relic whose abundance is set by mutual annihilations among multiple species.  Traditionally, this coannihilation scenario has been applied to weak scale dark matter that is highly degenerate with other states.  We show that coannihilation among states with split masses points to dark matter that is exponentially lighter than the weak scale, down to the keV scale.
  We highlight the regime where dark matter does not participate in the annihilations that dilute its number density. 
  In this ``sterile coannihilation" limit, the dark matter relic density is independent of its couplings, implying a broad parameter space of thermal relic targets for future experiments.
  Light dark matter from coannihilation evades stringent bounds from the cosmic microwave background, but will be tested by future direct detection, fixed target, and long-lived particle experiments.
\end{abstract}
\end{titlepage}

%%%%%%%%%%%%%%%%%%%%%%%%%%%%%%%%%%%%%%%%%%%%%%%%%%%%%%%%%%%%%%%%%%%%%%%%%%%%%%%%%%%%%%%%%%%

\tableofcontents
\setcounter{page}{2}

%%%%%%%%%%%%%%%%%%%%%%%%%%%%%%%%%%%%%%%%%%%%%%%%%%%%%%%%%%%%%%%%%%%%%%%%%%%%%%%%%%%%%%%%%%%
%%%%%%%%%%%%%%%%%%%%%%%%%%%%%%%%%%%%%%%%%%%%%%%%%%%%%%%%%%%%%%%%%%%%%%%%%%%%%%%%%%%%%%%%%%%
\section{Introduction} \label{sec:introduction}
Dark matter (DM) dominated the energy density of the Universe for much of its lifetime. Today it accounts for approximately a fifth of its energy budget.  Its microscopic origin is unknown, but several theoretical possibilities have been identified. In this work we focus on thermal relics, whose present abundance is determined by the freeze-out of their interactions in the early Universe. This set of DM candidates has the attractive feature of having a relic density that is insensitive to initial conditions and is tied to potentially measurable couplings.

The standard paradigm for thermal relics is the so-called WIMP (Weakly Interacting Massive Particle) miracle~\cite{Lee:1977ua,Kolb:1990vq,Gondolo:1990dk,Jungman:1995df}. In this scenario DM is kept in chemical equilibrium with the Standard Model (SM) thermal bath through its $2\to 2$ annihilations. When the rate of its annihilations becomes slower than the expansion rate, DM acquires a non-zero chemical potential and its number density freezes-out, redshifting with the expansion of the Universe until today. The standard calculation points to DM with weak scale interactions
\be \label{eq:WIMP}
\frac{\Omega_{WIMP}}{\Omega_{DM}}\approx \frac{(20\;{\rm TeV})^{-2}}{\langle \sigma v\rangle}\, .
\ee
However, increasingly stringent bounds on WIMPs from direct detection experiments~\cite{Angloher:2015ewa,Agnese:2015nto,Akerib:2015rjg,Aprile:2017iyp,Cui:2017nnn}, and the lack of obvious physics beyond the SM at the Large Hadron Collider, motivate exploration beyond the WIMP\@.

It has long been appreciated~\cite{Griest:1990kh} that at least three ``exceptions" exist to the standard WIMP computation: (1) mutual annihilation of multiple species ({\it coannihilation}), (2) annihilations into heavier states ({\it forbidden channels})~\cite{D'Agnolo:2015koa,Delgado:2016umt}, and (3) annihilations near a pole in the cross section~\cite{Ibe:2008ye,Feng:2017drg}.  A fourth exception was recently identified where the DM abundance is set by inelastic scattering instead of annihilations ({\it coscattering})~\cite{DAgnolo:2017dbv, Garny:2017rxs, Garny:2018icg}.  These four exceptions retain most of the standard assumptions of the WIMP paradigm.  DM starts in thermal equilibrium, has its number density diluted through $2\to 2$ annihilations, and has the same temperature as SM photons at freeze-out.  

The above exceptions correspond to small deformations of the assumptions underlying the WIMP miracle, and therefore correspond to close relatives of the WIMP within the broader theory space of DM candidates.  However, the phenomenology of the above exceptions can differ dramatically from the WIMP\@.  
It was recently shown that forbidden channels and coscattering both naturally lead to DM that is exponentially lighter than the weak scale~\cite{D'Agnolo:2015koa, DAgnolo:2017dbv}. In this work we explore the best known among the above exceptions, {\it coannihilation}, and show that it also naturally leads to exponentially light DM\@. Other mechanisms leading to DM lighter than the weak scale are explored for example by Refs.~\cite{Carlson:1992fn,Boehm:2003hm,Finkbeiner:2007kk,Pospelov:2007mp,Feng:2008ya, Kaplan:2009ag, Falkowski:2011xh,Hochberg:2014dra,Bernal:2015bla,Kuflik:2015isi, Dror:2016rxc, Kopp:2016yji, Knapen:2017xzo,Falkowski:2017uya,DAgnolo:2015nbz}.

%%%%%
\begin{figure}[!t]
\begin{center}
\includegraphics[width=0.6\textwidth]{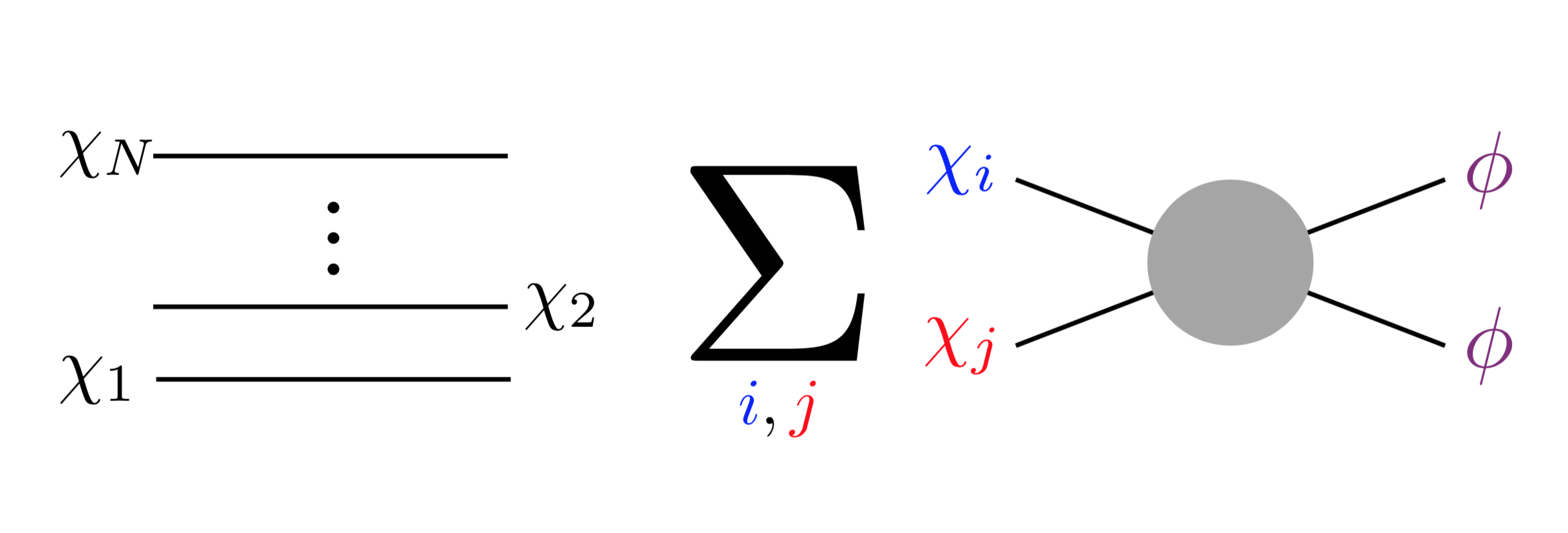}
\end{center}
\vspace{-.3cm}
\caption{Illustration of the collection of $N$ states, $\chi_i$, required to realize coannihilation ({\it left}) and the processes that control DM freeze-out ({\it right}).  DM annihilates into unstable states, $\phi$, from the thermal bath.  The $N$ states share a quantum number and eventually decay to the lightest state, $\chi_1$, which constitutes the DM today.  Freeze-out is determined by a summation over all possible annihilation modes among the collection of states.}
\label{fig:CoSchema}
\end{figure} 
%%%%%%

The dynamics of coannihilation are summarized in Fig.~\ref{fig:CoSchema}.  Multiple states are assumed to share the DM stabilizing symmetry, and they can be ordered by increasing mass (left of Fig.~\ref{fig:CoSchema}), $m_{\chi_1} < m_{\chi_2} <  \ldots < m_{\chi_N}$.  In general, freeze-out is determined by a summation of all possible annihilation channels (right of Fig.~\ref{fig:CoSchema}). We postpone a technical discussion of the relic density till later, but for now note that Ref.~\cite{Griest:1990kh} assumed that DM has a weak scale mass, and found that coannihilation is only relevant for weak scale DM when the various states are highly degenerate, with mass splittings below $\sim 1-10\%$.  Weak scale coannihilation has been studied extensively in supersymmetry when the lightest neutralino is degenerate with another superpartner~\cite{Edsjo:1997bg, Ellis:1998kh, Ellis:1999mm, Gomez:1999dk, Gomez:2000sj, Arnowitt:2001yh, Baer:2002fv,Mizuta:1992qp, Nihei:2002sc, ArkaniHamed:2006mb, Cheung:2012qy, BirkedalHansen:2001is, BirkedalHansen:2002am, Baer:2005zc, Baer:2005jq,Profumo:2004wk,Ellis:2001nx,Boehm:1999bj, Nagata:2015hha, Nagata:2015pra}, and also within non-supersymmetric models~\cite{Tulin:2012uq,Ibarra:2015nca,Baker:2015qna,Khoze:2017ixx,ElHedri:2017nny,Davoli:2017swj,Davoli:2018mau}. Here we explore coannihilation among multiple non-degenerate particles, and show that coannihilation opens up parameter space where DM can be orders of magnitude lighter than a WIMP\@. This was first pointed out in~\cite{ourtalks} and later studied for specific models~\cite{Izaguirre:2017bqb, Chauhan:2017eck}.

\begin{figure}[!t]
\begin{center}
\includegraphics[width=0.9\textwidth]{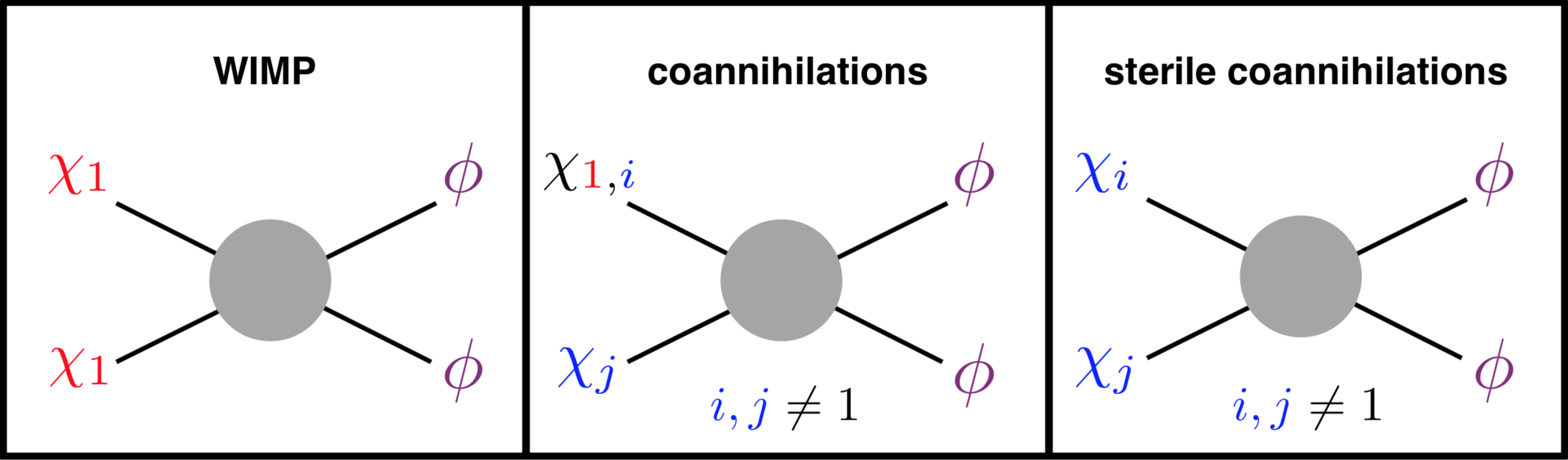}
\end{center}
\vspace{-.3cm}
\caption{Schematic representation of three phases of coannihilation for a set of DM particles, $\chi_i$, sharing a conserved quantum number. At late times, the heavier states decay to the lightest one, $\chi_1$, which comprises the DM today.
In the WIMP phase, $\chi_1 \chi_1$ annihilations dominate.  In the coannihilation phase, other annihilations dominate.  In the sterile coannihilation phase, $\chi_1$ does not participate in the dominant annihilations, and therefore the relic abundance of DM is independent of its couplings.
}
\label{fig:CoLayout}
\end{figure}

There are several phases for (co)annihilation, summarized in Fig.~\ref{fig:CoLayout}, depending on which type of annihilations dominates at freeze-out.  The left diagram shows the WIMP, where $\chi_1 \chi_1$ annihilations dominate.  The central diagram shows when coannihilation containing heavier particles in the set are important. The right diagram shows a special phase of coannihilation, that we dub {\it sterile coannihilation}, where the dominant annihilation modes do not contain DM in the initial state. Sterile coannihilation manifests in supersymmetry when the bino coannihilates with a wino~\cite{Baer:2005jq,ArkaniHamed:2006mb,Cheung:2012qy,Nagata:2015pra}, stop~\cite{Boehm:1999bj,Ellis:2001nx}, or gluino~\cite{Profumo:2004wk,Nagata:2015hha}, and in non-supersymmetric models~\cite{Ibarra:2015nca,Khoze:2017ixx,ElHedri:2017nny,Davoli:2018mau}, but we are the first to apply it to light DM\@. Sterile coannihilation is our main focus in this paper, and has the special property that the DM relic density is independent of the couplings of DM\@.  Experiments searching for light DM often focus on {\it thermal targets}~\cite{Battaglieri:2017aum}, where masses and couplings are chosen such that the DM matches the observed abundance, assuming a thermal relic cosmology.  It is usually stressed that thermal targets relate the DM mass to its couplings, but sterile coannihilation is a counterexample that points to a broader parameter space of viable models.

%%%%%
%%%%%%

\begin{figure}[!t]
\begin{center}
\includegraphics[width=0.6\textwidth]{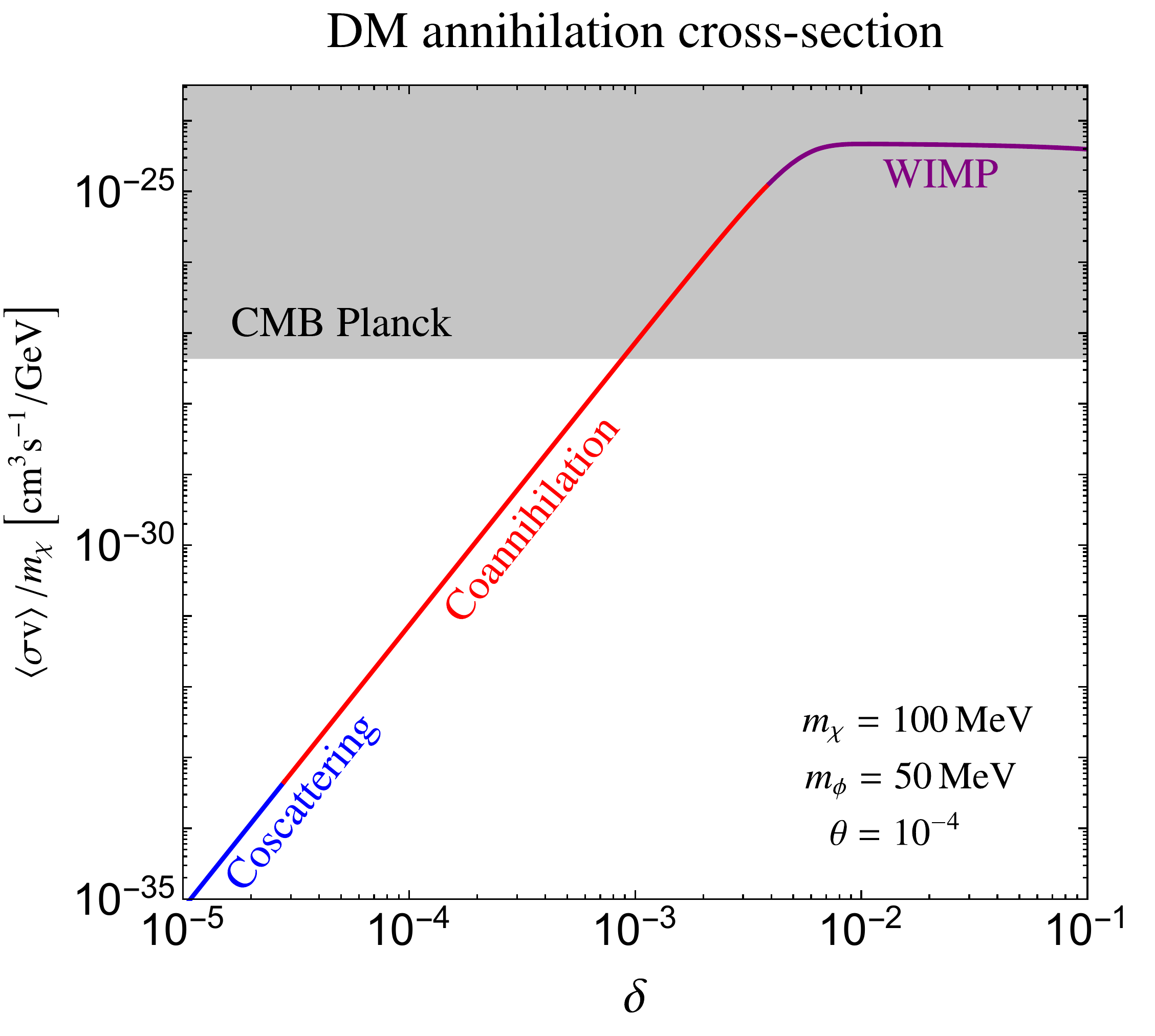}
\end{center}
\vspace{-.3cm}
\caption{The DM self-annihilation cross section, $\chi \chi \rightarrow \phi \phi$, as a function of the mixing angle, $\delta$, between mostly sterile DM, $\chi$, and an active state with rapid annihilations.  The DM mass is fixed to $m_\chi = 100$~MeV\@.  DM annihilates to a scalar, $\phi$, with mass $m_\phi=50~\mathrm{MeV}$ and mixing angle $\theta = 10^{-4}$ with the SM Higgs.   (See Sec.~\ref{sec:toyrelic} and~\ref{sec:secluded} for a detailed description of the model and definitions of these parameters.)  The color of the line delineates whether freeze-out is described by the WIMP (purple), coannihilation (red), or coscattering~\cite{DAgnolo:2017dbv} (blue) phase.  The gray region shows the Planck bound on DM annihilations~\cite{Ade:2015xua}, which excludes the WIMP phase for this DM mass.}
\label{fig:CMB}
\end{figure}

Coannihilation naturally evades stringent bounds from the Cosmic Microwave Background (CMB)~\cite{Ade:2015xua} because, while all annihilations contribute at freeze-out, the heavier states will typically decay down to  $\chi_1$ before recombination~\cite{Izaguirre:2014dua,ourtalks}.  Therefore, the CMB is only sensitive to $\chi_1 \chi_1$ annihilations, which may be suppressed if other annihilations dominate at freeze-out.  This is illustrated in Fig.~\ref{fig:CMB}, which shows the $\chi_1 \chi_1$ annihilation cross section, for a particular model that we introduce below, as a function of the mixing angle between DM, which begins with no interactions, and an active state that experiences rapid annihilations.  At large mixing, DM is WIMP-like and excluded by the CMB, whereas at small mixing coannihilation (or coscattering) sets the relic density and the CMB bound is evaded.

%%%%%
%%%%%%

The remainder of this paper is organized as follows.  In Sec.~\ref{sec:relic}, we review the formalism of Ref.~\cite{Griest:1990kh} for computing the relic density from coannihilation and we explain why coannihilation among multiple non-degenerate species naturally leads to DM that is exponentially lighter than the weak scale.  In Sec.~\ref{sec:toyrelic}, we introduce a toy model, with annihilations into dark states, that exhibits sterile coannihilation. In Sec.~\ref{sec:dark}, we complete the toy model by coupling it to either the SM or a dark sector thermal bath, and we study the detailed phenomenology and experimental prospects.  In Sec.~\ref{sec:SM}, we consider an example model with direct annihilations to SM particles.  Sec.~\ref{sec:conclusions} contains our conclusions.

%%%%%%%%%%%%%%%%%%%%%%%%%%%%%%%%%%%%%%%%%%%%%%%%%%%%%%%%%%%%%%%%%%%%%%%%%%%%%%%%%%%%%%%%%%%
%%%%%%%%%%%%%%%%%%%%%%%%%%%%%%%%%%%%%%%%%%%%%%%%%%%%%%%%%%%%%%%%%%%%%%%%%%%%%%%%%%%%%%%%%%%
\section{Relic Density} \label{sec:relic}
In this section, we begin by reviewing the formalism of Ref.~\cite{Griest:1990kh} for treating coannihilation.  We then include a novel discussion of how non-degenerate coannihilation points to dark matter that is exponentially lighter than the weak scale.  

We assume that DM is stabilized by a $\mathcal{Z}_2$ symmetry, or a larger symmetry with a $\mathcal{Z}_2$  subgroup.  We consider a collection of $N$ particles, $\chi_i$, depicted in Fig.~\ref{fig:CoSchema}, that include DM and have the same charge as DM under this stabilizing symmetry. We label them in such a way that $m_{\chi_i} < m_{\chi_j}$ for $i<j$.  We take $\phi$ to stand for any other state that interacts with $\chi_i$ and is not charged under the $\mathcal{Z}_2$ symmetry.  The relevant  processes in the DM Boltzmann equations are: annihilations $\chi_i \chi_j \to \phi \phi$, inelastic scatterings  $\chi_i \phi \to \chi_j \phi$, decays $\chi_j \to \chi_i \phi$, and the inverses of each of these processes. If the decays are sufficiently rapid, DM today is entirely composed of the lightest member of the set, $\chi_1$. We let $n_i$ refer to the number density of state $\chi_i$.  The DM relic abundance is determined by the total number density, $n\equiv \sum_i n_i$, at freeze-out. 

There are {\it a priori} $N$ separate Boltzmann equations, one for each of the $\chi_i$, but Ref.~\cite{Griest:1990kh} noticed an important simplification.  As long as scatterings or decays are rapid enough to maintain chemical equilibrium among the different states, $\chi_i \leftrightarrow \chi_j$, the ratios of abundances track their equilibrium values, $n_i / n \approx n_i^{\rm eq}/ n^{\rm eq}$, where $n_i^{\rm eq}$ and $n^{\rm eq}$ correspond to the equilibrium Boltzmann distributions.  This means that $N-1$ Boltzmann equations can be removed, and a single Boltzmann equation governs the evolution of $n$.  Coannihilation refers to this phase where scatterings and decays maintain chemical equilibrium until long after annihilations decouple, which is generically the case when $m_\phi \ll m_{\chi_1}$ and couplings are large enough.  For massive $\phi$ or small enough couplings, exchange among the $\chi_i$ can decouple before annihilations, leading to the coscattering phase~\cite{DAgnolo:2017dbv}.

In the coannihilation limit, where  $\chi_i \leftrightarrow \chi_j$ is rapid until after annihilations decouple, the evolution of $n$ in the early Universe is governed by the single Boltzmann equation
\be
\frac{dn}{dt}=-3 Hn-\sum_{i,j=1}^N\langle \sigma_{ij}v\rangle\left(n_i n_j -n_i^{\rm eq} n_j^{\rm eq}\right)\, , \label{eq:n}
\ee
where $\langle \sigma_{ij}v\rangle$ are the thermally averaged annihilation cross sections (see Fig.~\ref{fig:CoSchema}). Eq.~\ref{eq:n} can be further simplified to
\be \label{eq:BE}
\frac{dn}{dt}=-3 Hn-\langle \sigma_{\rm eff} v\rangle\left(n^2 -(n^{\rm eq})^2\right)\, ,
\ee
where we used $n_i / n \approx n_i^{\rm eq} / n^{\rm eq}$ and have defined an effective annihilation rate,
\be 
\sigma_{\rm eff} = \sum_{i,j=1}^N\sigma_{ij}\frac{n^{\rm eq}_i n^{\rm eq}_j}{\left(n^{\rm eq}\right)^2} \approx \sum_{i,j=1}^N\sigma_{ij}\frac{g_i g_j}{g_{\rm eff}^2}\left(1+\Delta_i\right)^{3/2}\left(1+\Delta_j\right)^{3/2}e^{-x(\Delta_i+\Delta_j)}\, , \label{eq:relicco}
\ee
where $x = m_{\chi_1} / T$, $\Delta_i \equiv \frac{m_{\chi_i} - m_{\chi_1}}{m_{\chi_1}}$, and $g_i$ counts the number of internal degrees of freedom. The last step in Eq.~\ref{eq:relicco} holds when all $\chi_i$'s are non-relativistic. As in~\cite{Griest:1990kh}, we find it convenient to define,
\be
g_{\rm eff}\equiv\sum_{i=1}^N g_i \left(1+\Delta_i\right)^{3/2}e^{-x\Delta_i}\, .
\ee

Eq.~\ref{eq:BE} is the same as the Boltzmann equation for a single WIMP that leads to Eq.~\ref{eq:WIMP}, except the WIMP annihilation rate is replaced by the total effective annihilation rate.  This implies that $\left< \sigma_{\rm eff} v \right>$ should be weak scale for coannihilation to reproduce the observed DM abundance.  We see that heavy states decouple exponentially from the effective annihilation rate, because the $ij$ term is suppressed by the factor $e^{- x (\Delta_i + \Delta_j)}$.  Ref.~\cite{Griest:1990kh}, and most of the following literature on coannihilation, have assumed that $\chi_i$ are weak scale particles and therefore that $\sigma_{ij}$ are weak scale.  In order to prevent the exponential suppression from being too sizable, $\Delta_i \ll 1$ is required.  This observation has led to the lore that coannihilation is only relevant among highly degenerate states.  This is the most widely studied regime of coannihilation that found multiple applications in supersymmetric phenomenology. In a classic example the DM particles are a combination of nearly degenerate bino and wino or higgsino,  and $\phi$ represents any light SM state, such as the electron~\cite{Edsjo:1997bg, ArkaniHamed:2006mb, Cheung:2012qy, BirkedalHansen:2001is,BirkedalHansen:2002am, Baer:2005zc,Baer:2005jq}.

In this work we highlight a different regime, where non-degenerate coannihilation leads to light DM~\cite{ourtalks}.  We assume that $\sigma_{11}$ is suppressed, so that $\sigma_{1i}$ or $\sigma_{ij}$ dominates, with $i,j \ne 1$ (see the right panel of Fig.~\ref{fig:CoLayout}).  We allow non-degeneracy: $\Delta_i \sim \mathcal{O}(1)$.  In order for $\sigma_{\rm eff}$ to be weak scale despite the exponential suppression, we require $\sigma_{ij}$ to be exponentially larger than the weak scale.  This requires a mass scale for the dark states that is exponentially lighter than the weak scale, $m_{\chi_i} \ll \mathrm{TeV}$.

We will focus in particular on the limit of {\it sterile coannihilation} (right of Fig.~\ref{fig:CoLayout}).  We imagine that the lightest particle in the DM family contributes negligibly to the total annihilation cross section: 
\be
\sigma_{11}, \sigma_{1i} \ll \sigma_{\rm eff}.
\ee
Sterile coannihilation leads to the interesting simplification that the DM abundance depends on the DM mass, $m_{\chi_1}$ but is independent of the DM couplings and only depends on the annihilation rate of the heavier states and the mass splittings.  At first it might seem surprising that annihilations that do not involve DM can deplete its abundance, but scattering and inverse decays allow $\chi_1$ to convert into heavier states, $\chi_1 \rightarrow \chi_i$, such that the annihilations of heavy states effectively remove $\chi_1$ too.

To highlight the parametrics of sterile coannihilation, we make the further assumption that the heaviest state is the most strongly interacting and that the other annihilation channels can be neglected. Then the total cross section simplifies to
\be
\sigma_{\rm eff} \approx \frac{g_N^2}{g_1^2}\sigma_{NN}\left(1+\Delta_N\right)^3e^{- 2 \Delta_N x}\, ,
\ee 
and the relic density acquires an exponential dependence on the mass splitting $\Delta_N$,
\be
\Omega_\chi h^2 \propto \frac{1}{\langle \sigma_{\rm eff} v\rangle_f}\propto e^{2 \Delta_N x_f}\, , \label{eq:relicpar}
\ee
where $x_f = m_{\chi_1} / T_f$ and $T_f$ is the temperature at which the heavy annihilations freeze-out.
The physical origin of this exponential is simple to understand. We have assumed that all the particles in the set are in equilibrium with each other (through scatterings or decays) until long after annihilations decouple. Therefore the chemical potential for $n$ is zero until the last annihilation process decouples from the thermal bath.  In our example the last particle to decouple is also the heaviest. So freeze-out occurs earlier than it would if only the lightest state was present with the same interactions, $T_f=m_N/x_f > m_1/x_f$, implying an exponentially enhanced relic density for DM\@.  The exponential in Eq.~\ref{eq:relicpar} can be very large for $\mathcal{O}(1)$ mass splittings, allowing for dramatic departures from the standard WIMP lore.

If we fix the relic density to its observed value, in the sudden freeze-out approximation, $x_f$ is exactly the same as that of a WIMP and depends only on the DM mass. This is because for all thermal relics, the observed DM energy density today is,
\be \label{eq:universal}
\rho_{DM}=m_{DM} n_{DM} (x_f) \frac{s_0}{s_f}, 
\ee
where $s_0$ and  $s_f$ are the entropy densities today and at freeze-out, respectively.  Eq.~\ref{eq:universal} does not depend on the dynamics of freeze-out and is therefore universal for thermal relics, and allows $x_f$ to be solved as a function of $m_{DM}$.
Typical values of $x_f$ for thermal relics with weak scale masses are around 25 going down to about 10 at 10~keV, so $e^{-\Delta_i x_f}$ can be a large exponential suppression in sterile coannihilation.

We can also study $x_f$ without fixing the relic density, using the sudden freeze-out approximation, to see how $x_f$ depends on parameters such as the cross section and masses. In coannihilation freeze-out occurs when $n \langle \sigma_{\rm eff} v\rangle \approx H$. If $\Delta_i \approx \mathcal{O}(1)$, when DM is non-relativistic $n\approx n_1$. We can use the sudden freeze-out approximation to solve for $x_f$ in the usual way, with the only difference coming from the exponentials in $\sigma_{\rm eff}$. In the previous example, where the $N$-th state dominates, we have (for $s$-wave annihilations)
\be
n_1 \langle \sigma_{\rm eff} v\rangle \approx \frac{g_N^2 m_1^3 \left(1+\Delta_N\right)^3}{g_1(2\pi x)^{3/2}}\sigma_{NN}e^{-(1+2\Delta_N)x}\, .
\ee
So $n_1 \langle \sigma_{\rm eff} v\rangle_f \approx H_f$ gives the right relic density at
\be
x_f \approx  \frac{1}{1+2\Delta_N } \left( 21+\log\left[\frac{g_N^2\left(1+\Delta_N\right)^3m_{1}\sigma_{NN}}{g_1g_{*}^{1/2}\text{GeV}\times\text{pb}}\right]+\frac{1}{2}\log x_{f} \right) \, ,
\ee
where $M_{\text{Pl}}$ is the Planck mass and $g_{*}$ is the total number of relativistic degrees of freedom at freeze-out.  As for the WIMP case, there is logarithmic dependence on the cross section and absolute mass scale, but unlike the case of the WIMP, there is also non-logarithmic dependence on the mass splitting through the factor $(1+2\Delta_N)^{-1}$.

The numerical results that follow use \texttt{MicrOMEGASv4}~\cite{Barducci:2016pcb} to compute the relic density, unless otherwise specified.  We have cross-checked \texttt{MicrOMEGASv4} with our own numerical solutions of the Boltzmann equation and find good agreement.

%%%%%
\begin{figure}[!t]
\begin{center}
\includegraphics[width=0.9\textwidth]{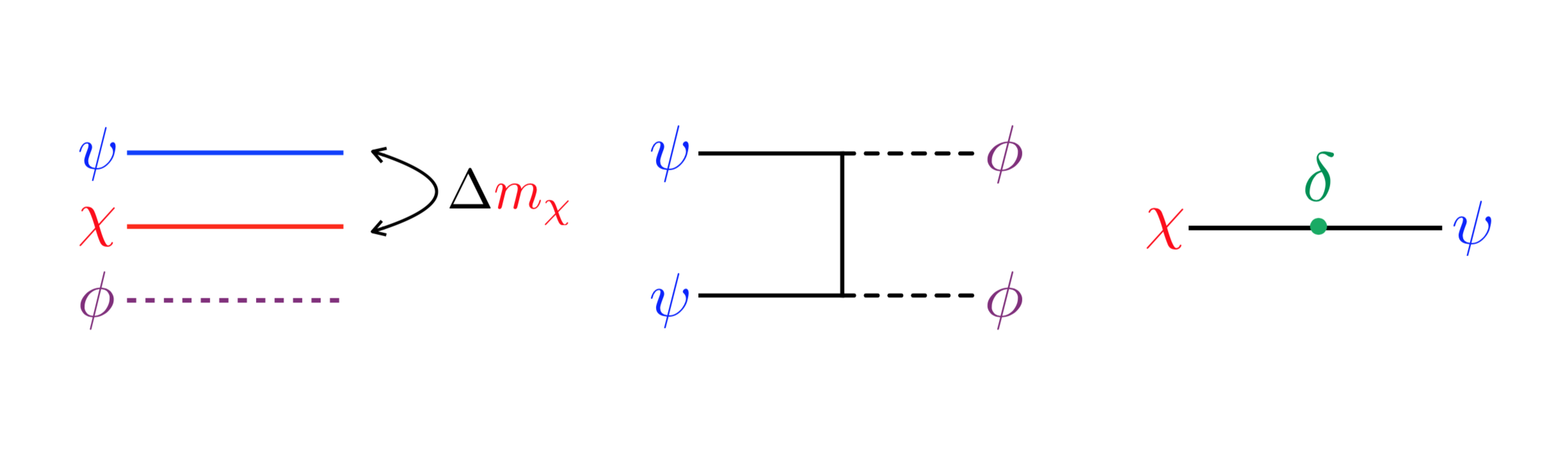}
\end{center}
\vspace{-.75cm}
\caption{Schematic representation of the model described in Sec.~\ref{sec:toyrelic}. From left to right: the mass spectrum, main annihilation channel, and strength of the mixing between DM, $\chi$, and its heavier coannihilating partner, $\psi$.}
\label{fig:ToyLayout}
\end{figure} 
%%%%%%

%%%%%
\begin{figure}[!t]
\begin{center}
\includegraphics[width=0.9\textwidth]{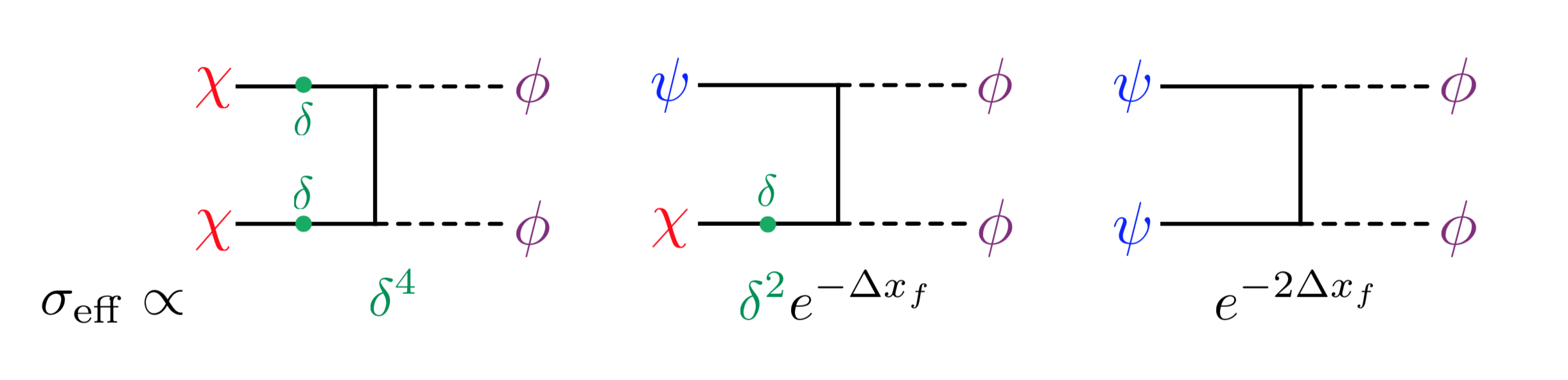}
\end{center}
\vspace{-.3cm}
\caption{Annihilation channels for the model presented in Sec.~\ref{sec:toyrelic}. The DM particle can approximately be identified with $\chi$. In the regime: $\delta^2 \ll e^{-\Delta x_f}$, $\psi\psi$ annihilations dominate and we realize sterile coannihilation.  In the opposite limit, $\delta^2 \gg e^{-\Delta x_f}$, $\chi \chi$ annihilations dominate and the model becomes WIMP-like.}
\label{fig:ToyDiagrams}
\end{figure} 
%%%%%%

%%%%%%%%%%%%%%%%%%%%%%%%%%%%%%%%%%%%%%%%%%%%%%%%%%%%%%%%%%%%%%%%%%%%%%%%%%%%%%%%%%%%%%%%%%%
\subsection{The Relic Density in a Simple Model}\label{sec:toyrelic}
To make some of these ideas more concrete we introduce a simple model where sterile coannihilation is realized. This model will serve as the main building block for our discussion of coannihilation in dark sectors weakly coupled to (or decoupled from) the SM\@.  It is a toy model (in this section) in the sense that we do not specify a coupling to radiation, which can take different forms and will determine the phenomenology of the theory, as we discuss in the next section. 

We consider a Majorana fermion, $\psi$, which will act as the heavy particle that dominates the annihilation rate.  We take $\psi$ to have rapid annihilations into a scalar particle, $\phi$, $\psi \psi \rightarrow \phi \phi$.  Then we add DM in the form of a Majorana fermion, $\chi$, which begins as an inert particle and inherits interactions through a small mass-mixing with $\psi$.  In the small mixing limit, annihilations with $\chi$ in the initial state are suppressed, so that $\psi \psi$ annihilations naturally dominate.  This setup is achieved by the following potential,
\be
V= \frac{m_\phi^2}{2}\phi^2+\frac{m_\psi}{2}\psi^2+\frac{m_\chi}{2}\chi^2+\delta m \chi \psi + \frac{y}{2} \phi \psi^2 +\mathrm{h.c.}\,  \, .
\label{eq:toyV}
\ee
There are two physical phases in $V$ that we choose to parametrize by allowing $\delta$ and $y$ to be complex. Notice that for generic phases of these parameters, the relevant annihilation processes are {\it s}-wave. We take $|\delta|\equiv |\delta m/m_\chi| \ll 1$ so $\chi$ is mostly sterile and $m_\psi \gtrsim m_\chi > m_\phi$. This choice of parameters realizes in a simple way our sterile coannihilation scenario with $\chi_1\approx\chi$ and $\chi_2\approx\psi$. This is shown schematically in Fig.~\ref{fig:ToyLayout}, where, going from left to right, we sketch the mass spectrum, main annihilation channel, and strength of the mixing between DM and its coannihilating partner.  

We assume that $\phi$ has a small coupling with the SM or with an additional light dark state, sufficient to keep the DM in kinetic and chemical equilibrium with radiation at freeze-out, but small enough that annihilations of $\chi$ and $\psi$ to $\phi\phi$ dominate over direct annihilations to lighter states.

%%%%%
\begin{figure}[!t]
\begin{center}
\includegraphics[width=\textwidth]{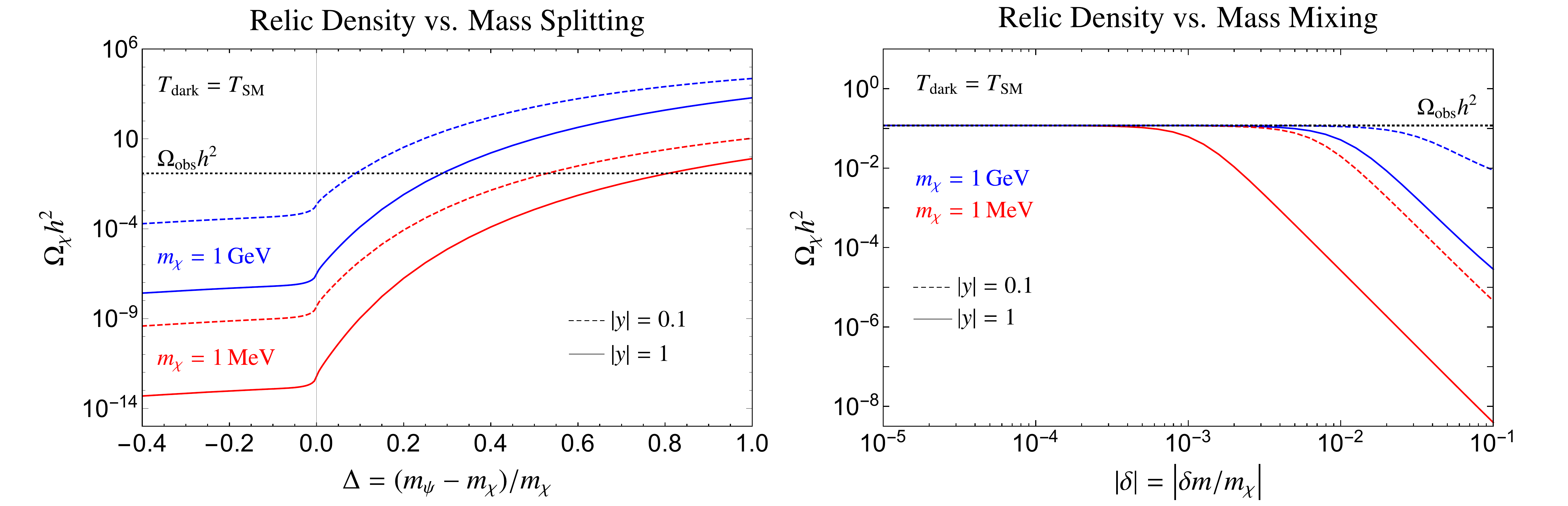}
\end{center}
\vspace{-.3cm}
\caption{{\it Left}: Dependence of the relic density on the mass splitting $\Delta$ between DM, $\chi$, and its rapidly annihilating partner, $\psi$. For $\Delta >0$ the relic density grows exponentially with the mass splitting, such that it crosses the observed value, $\Omega_{\rm obs} h^2 \approx 0.12$, for a particular value of the splitting.  {\it Right}: Dependence of the relic density on the mass mixing, $\delta$, between DM and its partner, $\psi$. At large $\delta$ we exit the sterile coannihilation regime discussed in the text and lose the exponential dependence of the relic density on the mass splitting $\Delta$. Both figures were made for generic values of the phases:  $y=|y|e^{i \pi/4}$ and $\delta=|\delta|e^{i \pi/4}$.}
\label{fig:OmegavsDelta}
\end{figure} 
%%%%%%

In Fig.~\ref{fig:ToyDiagrams}, we show the three possible types of annihilations: $\chi \chi, \chi \psi$, and $\psi \psi$.  When computing the effective annihilation rate, annihilations are suppressed by a factor of $\delta^2$ for each $\chi$ in the initial state, and by $e^{- x \Delta}$ for each $\psi$ in the initial state, where $\Delta \equiv (m_\psi-m_\chi)/m_\chi$.  Therefore, sterile coannihilation ($\psi \psi$ dominating) is realized in the limit $\delta^2 \ll e^{-\Delta x_f}$.  In the opposite limit $\delta^2 \gg e^{-\Delta x_f}$, the model becomes WIMP-like, with  $\chi \chi$ annihilations dominating.

\begin{figure}[!t]
\begin{center}
\includegraphics[width=\textwidth]{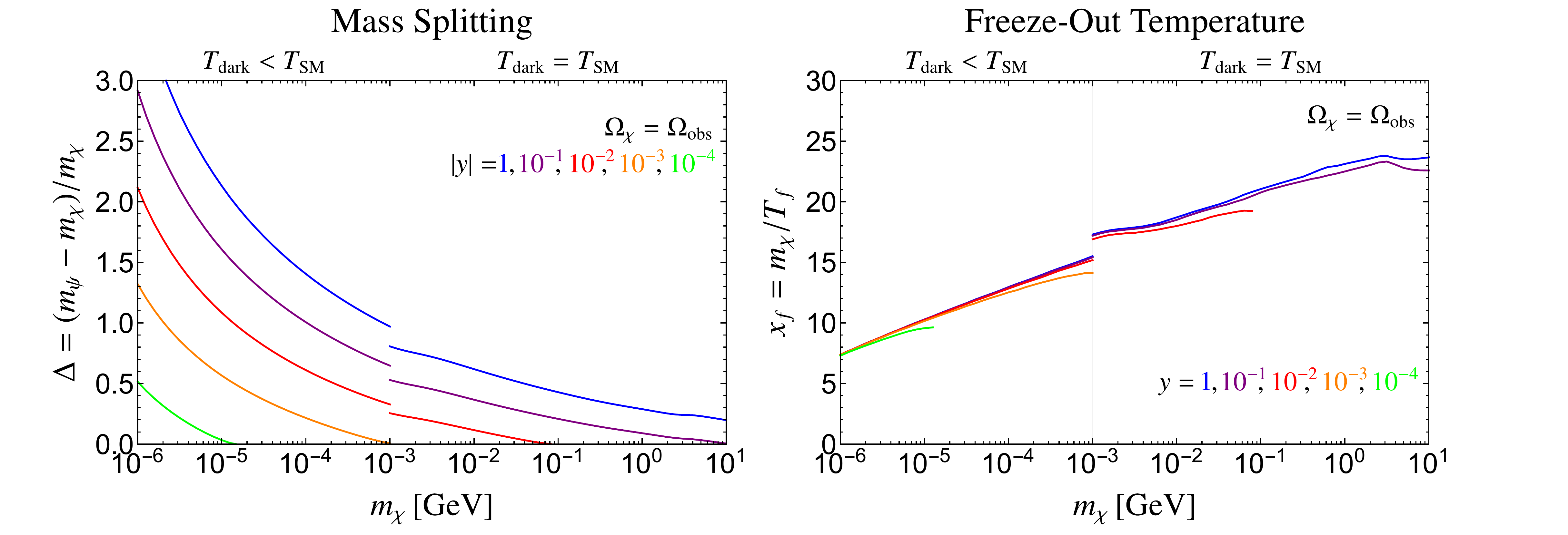}
\end{center}
\vspace{-.3cm}
\caption{{\it Left}: Value of the mass splitting, $\Delta$, that gives the observed relic density. We plot $\Delta$ as a function of $m_\chi$ and the coupling, $y$, controlling the size of the annihilation cross section. Light dark sectors require generic, $\mathcal{O}(1)$, mass splittings. {\it Right}: $x_f$ as a function of $m_\chi$ and the coupling $y$. $x_f$ in coannihilating dark sectors is the same as for a WIMP, after fixing the DM relic density to the observed value, and exhibits the same logarithmic dependence on the DM mass. Both figures were made for generic values of the phases: $y=|y|e^{i \pi/4}$ and $\delta=|\delta|e^{i \pi/4}$. For $m_\chi<$~MeV we show the results for a decoupled dark sector in equilibrium with its own dark radiation (see Sec.~\ref{sec:decoupled}).}
\label{fig:Delta&xf}
\end{figure}

In Fig.~\ref{fig:OmegavsDelta} we illustrate the main qualitative point made in the previous section. The relic density grows exponentially with $\Delta \equiv (m_\psi-m_\chi)/m_\chi$. In the sterile coannihilation limit, 
\bea
\langle \sigma_{\rm eff} v\rangle &\approx& \langle\sigma_{\psi\psi} v \left(1+\Delta\right)^3e^{- 2 \Delta x}\rangle
\nn \\ \label{eq:sigmapsi}
&=&\frac{y^2_r y^2_i (1+\Delta)^4}{2\pi m_\chi^2}\frac{\sqrt{(1+\Delta)^2-r^2}}{\left[2(1+\Delta)^2-r^2\right]^2}e^{- 2 \Delta x}+\mathcal{O}(v^2)+\mathcal{O}(\delta^2)\, ,
\eea
where $y_r \equiv{\rm Re}[y]$, $y_i \equiv {\rm Im}[y]$, and $r\equiv m_\phi/m_\chi$. If the mass scale of the dark sector decreases or the coupling $y$ increases, the total cross section becomes larger, requiring a larger $\Delta$ to get the right relic density. This is shown in the left panel of Fig.~\ref{fig:Delta&xf}. 

%%%%%
%%%%%%

In the same figure we show that $x_f$ does not differ considerably from a WIMP and, as usual, decreases logarithmically with the DM mass. The last parameter of interest is the mixing $\delta$ between the DM, $\chi$, and the active state $\psi$. If it is sufficiently small, the relic density does not depend on it, as suggested by Eq.~\ref{eq:sigmapsi}. Only when we exit the sterile coannihilation regime does $\delta$ start to  impact $\sigma_{\rm eff}$. This is depicted in the right panel of Fig.~\ref{fig:OmegavsDelta}. 

In Fig.~\ref{fig:Delta&xf}, when $m_\chi$ is below 1~MeV, we imagine that the dark sector is completely decoupled from the SM\@. (As we discuss below, limits from $N_{\rm eff}$ exclude DM masses lighter than the MeV scale when DM is in kinetic equilibrium with the SM\@.)  In this case we add a light fermion species, with a sub-eV mass and  coupling to $\phi$, that acts as dark radiation and insures the scaling $T_{\rm dark}\sim a^{-1}$.  In general, the SM and dark sectors evolve with separate temperatures, when they are kinetically decoupled, that are determined by the separate conservation of entropy in each sector\cite{Feng:2008mu}
\be
\frac{T_{\rm dark}}{T_{SM}} =\xi_R \left(\frac{g_{*S}^{SM}(T_{SM})}{g_{*S}^{SM}(T^R_{SM})}   \frac{g_{*S}^{\rm dark}(T^R_{\text{dark}})}{g_{*S}^{\rm dark}(T_{\rm dark})} \right)^{1/3} \, , \label{eq:entropy}
\ee
where $T^R_{\text{dark}, SM}$ are the initial dark and SM temperatures after reheating, respectively, and $\xi_R\equiv T^R_{\text{dark}}/T^R_{SM}$.  In Fig.~\ref{fig:Delta&xf}, we assume for simplicity that the dark and SM sectors reheat to the same temperature, $\xi_R = 1$, above the weak scale.  For our dark sector this leads to $T_{\rm dark}<T_{\rm SM}$ at freeze-out, such that current $N_{\rm eff}$ constraints are naturally evaded. We defer a more detailed discussion of decoupled dark sectors to Sec.~\ref{sec:decoupled}.

%%%%%%%%%%%%%%%%%%%%%%%%%%%%%%%%%%%%%%%%%%%%%%%%%%%%%%%%%%%%%%%%%%%%%%%%%%%%%%%%%%%%%%%%%%%
%%%%%%%%%%%%%%%%%%%%%%%%%%%%%%%%%%%%%%%%%%%%%%%%%%%%%%%%%%%%%%%%%%%%%%%%%%%%%%%%%%%%%%%%%%%
\section{Annihilations to Dark Sector Particles} \label{sec:dark}

In this section we complete the toy model of  Sec.~\ref{sec:toyrelic} by explicitly introducing a coupling to radiation. We require the coupling to be large enough that DM is in kinetic equilibrium with the radiation until after freeze-out, which ensures that $T_{\rm dark}\sim a^{-1}$.  This scaling was assumed when deriving the relic density in Sec.~\ref{sec:relic}.  If we relax the assumption that DM is in thermal contact with radiation, then the dark sector will undergo a phase of cannibalism~\cite{Carlson:1992fn, deLaix:1995vi, Yamanaka:2014pva, Bernal:2015ova, Bernal:2015xba, Kuflik:2015isi, Soni:2016gzf, Pappadopulo:2016pkp, Farina:2016llk}, where $T_{\rm dark}$ drops only logarithmically with the scale factor.  We leave the study of coannihilation in a cannibalizing sector for future work.

There are two distinct possibilities for how DM may couple to radiation. The dark sector may couple directly to the SM, or the dark sector may contain its own light degrees of freedom that act as dark radiation. In the first case, constraints on $N_{\rm eff}$ do not allow for DM lighter than the MeV scale, while a completely decoupled dark sector can have DM as light as the keV scale.  As we discuss below, sub-keV masses are excluded by limits on warm DM from Lyman-$\alpha$ measurements.

We consider a dark sector that is the same as the toy model introduced in the previous section, for both the coupling to SM and dark radiation.  We have two Majorana fermions $\chi$ and $\psi$, and one real scalar $\phi$ interacting through the potential of Eq.~\ref{eq:toyV}. We take $m_\psi \gtrsim m_\chi > m_\phi$, with $\chi$ very weakly interacting through mixing of size $|\delta|= |\delta m/m_\chi| \ll 1$.  As above we define $\Delta\equiv (m_\psi - m_\chi)/m_\chi$.  Because we will always take the limit of very small mixing, $\delta \ll 1$, the mass eigenstates are approximately flavor eigenstates: $\chi_1 \approx \chi$ and $\chi_2 \approx \psi$.  In the following we will abuse notation, for simplicity, and refer to the mass eigenstates also as $\chi$ and $\psi$.

\subsection{Higgs Portal}\label{sec:secluded}
%%%%%
\begin{figure}[!t]
\begin{center}
\includegraphics[width=0.9\textwidth]{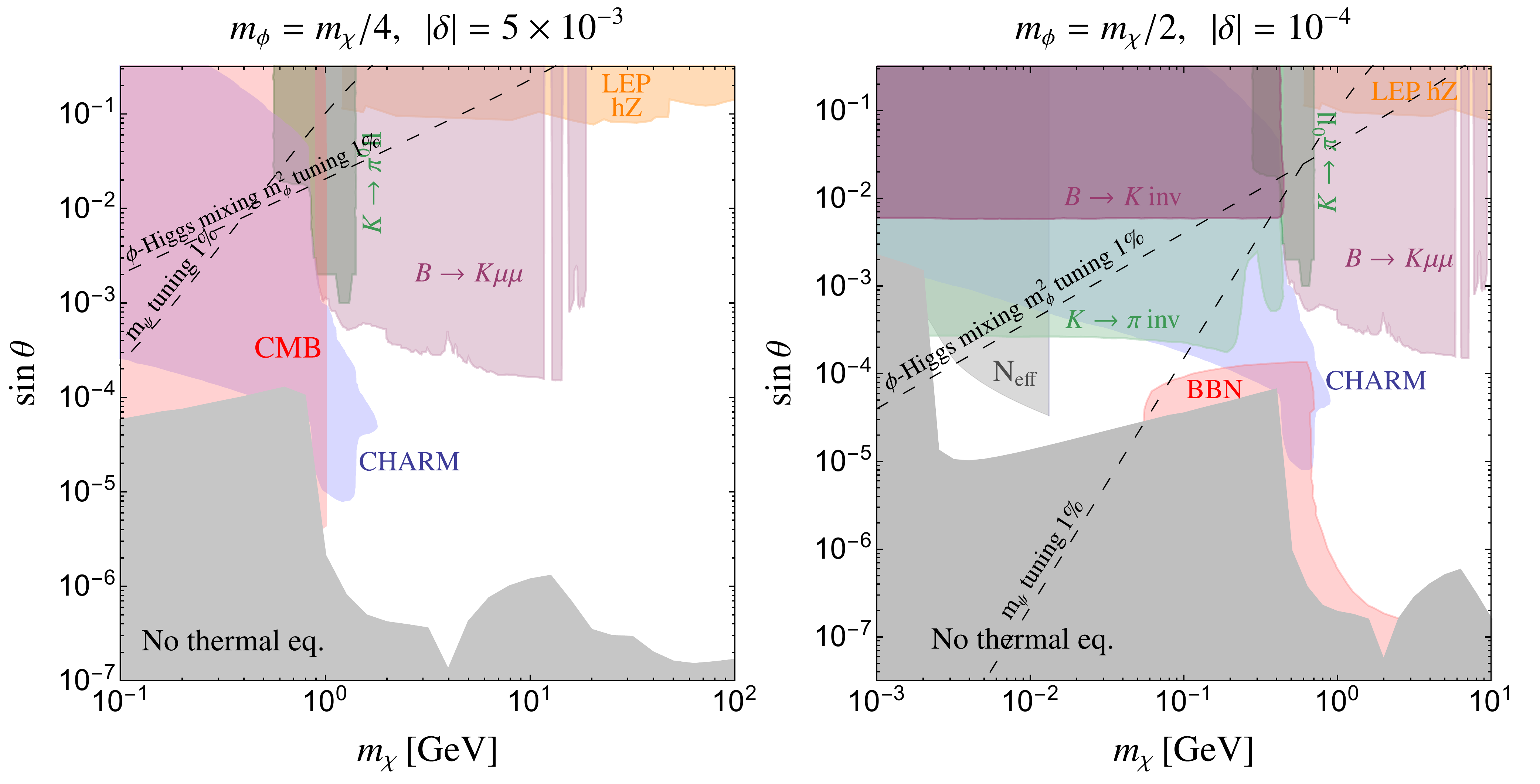}
\end{center}
\vspace{-.3cm}
\caption{Phenomenology of a dark sector weakly coupled to the SM\@. Both panels show current constraints as a function of the DM mass, $m_\chi$, and the size of the mixing between the dark sector scalar and the SM Higgs boson, $\sin \theta$. 
The {\it left} panel shows the effect of a lighter mediator, $\phi$, and larger DM mixing, $\delta$, compared to the heavier and more weakly coupled case in the {\it right} panel. The gray shaded area at the bottom of the plot covers values of the mixing for which the dark sector is not in equilibrium with the SM at freeze-out. The other bounds are discussed in the text and come from CMB measurements from Planck~\cite{Ade:2015xua}, BBN~\cite{Jedamzik:2006xz, Poulin:2015opa}, meson decays~\cite{AlaviHarati:2000hs,AlaviHarati:2003mr,Adler:2004hp,Chen:2007zk,Artamonov:2009sz,delAmoSanchez:2010bk,Aaij:2015tna,Aaij:2016qsm}, CHARM~\cite{Bergsma:1985qz}, and LEP~\cite{Acciarri:1996um,Schael:2006cr}. In every point of the plot $\Delta$ is fixed to reproduce the observed relic density. The remaining parameters are set to  $y=e^{i \pi/4}$, $m_\phi = m_\chi/4\ (m_{\chi}/2)$, and  $\delta = 5\times 10^{-3}e^{i \pi/4} \ (10^{-4}e^{i \pi/4} )$ in the {\it left} ({\it right}) panel.
}
\label{fig:phenoHiggs}
\end{figure} 
%%%%%%

%%%%%
\begin{figure}[!t]
\begin{center}
\includegraphics[width=0.9\textwidth]{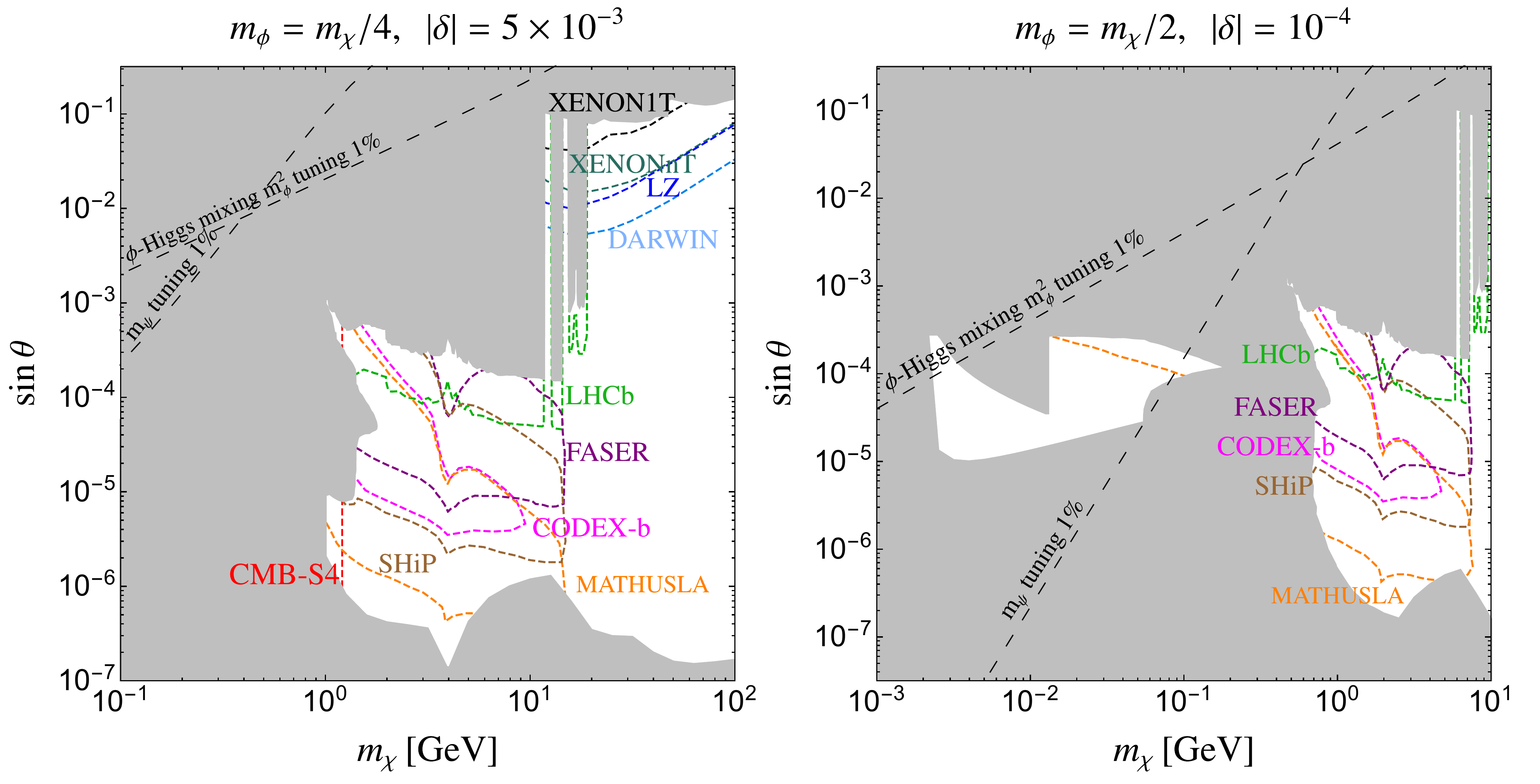}
\end{center}
\vspace{-.3cm}
\caption{Phenomenology of a dark sector weakly coupled to the SM\@. Both panels show the projected reach of future experiments as a function of the approximate DM mass, $m_\chi$, and the mixing between the dark sector scalar and the SM Higgs boson, $\sin \theta$. 
 The gray shaded area summarizes current constraints, presented separately in Fig.~\ref{fig:phenoHiggs}. The reach for direct detection experiments~\cite{Aprile:2012zx, Aprile:2015uzo, Aalbers:2016jon, Mount:2017qzi}, LHC-b~\cite{Gligorov:2017nwh}, CODEX-b~\cite{Gligorov:2017nwh} as derived in~\cite{Evans:2017kti}, FASER~\cite{Feng:2017uoz,Feng:2017vli}, MATHUSLA~\cite{Chou:2016lxi,Evans:2017lvd}, SHiP~\cite{Alekhin:2015byh,Anelli:2015pba} as derived in~\cite{Evans:2017lvd}, and CMB Stage-4 (CMS-S4)~\cite{Abazajian:2016yjj} is shown. In every point of the plot $\Delta$ is fixed to reproduce the observed relic density. The remaining parameters are set to  $y=e^{i \pi/4}$, $m_\phi = m_\chi/4\ (m_{\chi}/2)$, and $\delta = 5\times 10^{-3}e^{i \pi/4} \ (10^{-4}e^{i \pi/4} )$ in the {\it left} ({\it right}) panel.}
\label{fig:phenoHiggs2}
\end{figure} 
%%%%%%

We first consider the case in which the dark sector couples to the SM through the relevant Higgs portal coupling $a_\phi \phi |H|^2$,
\be
V\supset a_\phi \phi |H|^2+\frac{\mu_\phi^2}{2}\phi^2+\frac{\lambda_{\phi}}{4!}\phi^4-\frac{m_H^2}{2} |H|^2+\frac{\lambda}{4}|H|^4\, .
\ee
After electroweak symmetry breaking, we can parametrize the strength of the portal coupling using the mixing angle between $\phi$ and the SM Higgs boson
\be
\tan 2 \theta = \frac{4 a_\phi v}{\lambda v^2-\lambda_{\phi}v_\phi^2-2\mu_\phi^2}\, ,
\ee
where $v\approx 246$~GeV and $v_\phi\equiv\left<\phi\right>$ is the VEV of $\phi$. 

The decay width and branching fractions of $\phi$ are crucial for determining both the experimental prospects and when the SM and dark sector are in thermal equilibrium.
We take the scalar $\phi$ to be lighter than DM, such that it can only decay to SM particles.  We note that the width and branching fractions  of $\phi$ cannot be determined with precision when its mass is between $\sim 2m_\pi$ and a few GeV, due to significant hadronic uncertainties~\cite{Clarke:2013aya}.  For the width and branching fractions, we follow Ref.~\cite{Feng:2017vli}.

The phenomenology of the model is summarized in Figs.~\ref{fig:phenoHiggs} and~\ref{fig:phenoHiggs2}.  These figures show $m_\chi$ versus the singlet-Higgs mixing, $\sin \theta$, with $\Delta$ chosen at each point so that the relic density matches observation.  Fig.~\ref{fig:phenoHiggs} shows current constraints, while Fig.~\ref{fig:phenoHiggs2} presents the projected reach of future experiments. There is a lower bound on $\sin \theta$ set by the requirement that $\phi$ interactions with the SM are efficient at keeping the two sectors in thermal equilibrium at DM freeze-out.  Within the dark gray shaded areas at the bottom of Fig.~\ref{fig:phenoHiggs}, the two sectors are not in equilibrium when DM freezes-out.  Thermalization can be achieved by rapid $\phi$ decay or $2\rightarrow 2$ scattering processes. We find that $\phi$ decay is more efficient for our choice of parameters. 
The CMB constraints on $N_{\rm eff}$ and on the DM annihilation cross section~\cite{Ade:2015xua} in Fig.~\ref{fig:phenoHiggs} set a lower bound on $m_\chi$, almost independent of the size of the $\phi$-Higgs mixing angle. These two constraints are common to all realizations of coannihilation coupled to the SM, and we now describe them in more detail.

The CMB bound on $N_{\rm eff}$ excludes $\chi$ masses below $\sim10$ MeV\@.  If the dark sector is in equilibrium with photons after neutrinos decouple, then DM transfers entropy to the photon bath, increasing the photon temperature relative to neutrinos~\cite{Boehm:2013jpa}.  This is a familiar effect in the SM, where photons are heated relative to neutrinos after electrons turn non-relativistic.  Here, DM {\it lowers} $N_{\rm eff}$, since neutrinos are cooled relative to the photon temperature.  There is one important subtlety: if  $\sin \theta$ is sufficiently small, the dark sector may not be in kinetic equilibrium with the SM at neutrino decoupling, although it may later equilibrate before DM annihilations decouple~\cite{Berlin:2017ftj}. In this region, the value of $N_{\rm eff}$ probed by the CMB is sensitive to initial conditions.  We can always choose a low enough reheating temperature, compatible with BBN, for which there is no constraint on $N_{\rm eff}$. For this reason, we do not plot the $N_{\rm eff}$ bound in this part of the parameter space, as can be seen in the right panel of Fig.~\ref{fig:phenoHiggs}, where the light gray shaded region ends at $\sin \theta \approx 10^{-4} - 10^{-5}$.  We do not plot this constraint in the left panel of Fig.~\ref{fig:phenoHiggs} since it is everywhere weaker than the CMB bound on DM annihilations.

DM annihilations, $\chi\chi \rightarrow \phi\phi$, with subsequent $\phi$ decay to SM particles, can alter the recombination history through energy injection. This can lead to observable modifications to the CMB temperature and polarization power spectra~\cite{Chen:2003gz, Padmanabhan:2005es}. The rate of energy release per unit volume is proportional to $f_{\text{eff}}\langle \sigma v \rangle/m_{\rm DM}$, becoming larger for light thermal relics, which are excluded below about 10~GeV, depending on the annihilation products~\cite{Ade:2015xua, Slatyer:2015jla, Slatyer:2015kla}.   (A notable exception is DM annihilations to neutrinos, which are not strongly constrained by the CMB~\cite{Batell:2017cmf,Schmaltz:2017oov}.)  In our model, the strongly interacting heavier state, 
$\psi$, decays down to DM, $\chi$, before recombination.  The CMB is only sensitive to DM self-annihilations, which have cross section suppressed by small $\delta$.   Therefore, much lighter DM masses are compatible with CMB annihilations in our model compared to WIMPs.  The efficiency factor, $f_{\rm eff}$, measures the fraction of the DM mass that is converted to heating the photon-baryon plasma, and captures the model dependence of the DM annihilation products.
We follow the prescription from~\cite{Elor:2015bho} to calculate $f_{\text{eff}}$ for the cascade annihilation and use $f_{\text{eff}}$ tables provided in~\cite{Slatyer:2015jla}. The corresponding Planck bound~\cite{Ade:2015xua} is shown in the left panel of Fig.~\ref{fig:phenoHiggs} as a red shaded area. In Fig.~\ref{fig:phenoHiggs2}, we show the projected improvement of a factor of 3 in cross section reach from CMB Stage-4, assuming 60\% sky coverage~\cite{Abazajian:2016yjj}. In the right panel of Fig.~\ref{fig:phenoHiggs}, we choose a smaller value of $\delta$ where the bound is completely absent. 

An additional constraint comes from BBN and applies when $\phi$ is heavier, as in right panel of Fig.~\ref{fig:phenoHiggs} (red shaded area).  As $\psi$ becomes heavier, the splitting, $\Delta$, that matches the observed relic density becomes smaller. When the DM mass is greater than about 100 MeV, $\psi$ can only decay through an off-shell $\phi$ to $\chi$ plus SM states.  For small mixing angle $\theta$, the $\psi$ lifetime can become long enough to affect BBN through injection of electromagnetic particles into the primordial plasma~\cite{Jedamzik:2006xz, Poulin:2015opa}. When the mass splitting between $\psi$ and $\chi$ is larger than $2m_{\pi}$, the dominant decay channels have mesons in the final state.  However, we find that the strongest bound comes from the electromagnetic fraction of these decays, because the abundance of  $\psi$ is too small to affect proton-neutron conversions at early stages of BBN~\cite{Fradette:2017sdd}, and the injected hadrons are not energetic enough for hadro-dissociation to affect primordial light element abundances~\cite{Henning:2012rm,Kawasaki:2017bqm}.

A variety of other observations are also relevant to our parameter space. Constraints from meson decays~\cite{Wei:2009zv, Aaij:2015tna,Aaij:2016qsm,Schmidt-Hoberg:2013hba,Adler:2004hp,Artamonov:2009sz,  Chen:2007zk,delAmoSanchez:2010bk,Clarke:2013aya,AlaviHarati:2000hs,AlaviHarati:2003mr} and colliders~\cite{Acciarri:1996um,Schael:2006cr,Flacke:2016szy} follow directly from $\phi$'s interactions with the SM, and are unrelated to DM\@. In our model these constraints play a role at large mixing, $\sin \theta \gtrsim 10^{-3}$, as shown in Fig.~\ref{fig:phenoHiggs}. The CHARM proton beam dump searched for long-lived particles~\cite{Bergsma:1985qz}, and excludes part of the parameter space of a scalar mixed with the Higgs~\cite{Bezrukov:2009yw, Clarke:2013aya, Alekhin:2015byh}. In this case, the scalar $\phi$ can be produced from meson decays, and can be detected when it decays within the displaced detector.
 
The most promising avenues for probing more parameter space are DM direct detection~\cite{Aprile:2012zx, Mount:2017qzi, Aprile:2015uzo, Aalbers:2016jon}, as can be seen in the left panel of Fig.~\ref{fig:phenoHiggs2}, and future beam dump experiments such as SHiP~\cite{Alekhin:2015byh,Anelli:2015pba}. Experiments exploiting the abundance of mesons in LHC collisions, such as CODEX-b~\cite{Gligorov:2017nwh}, MATHUSLA~\cite{Chou:2016lxi, Evans:2017lvd}, and FASER~\cite{Feng:2017uoz, Feng:2017vli} are competitive with SHiP\@. In  Fig.~\ref{fig:phenoHiggs2} we also show the projected reach for LHCb as derived in Ref.~\cite{Gligorov:2017nwh}.
The reach of these experiments depends on the type of portal that connects the SM to the dark sector and the size of the portal coupling.  However, we note that any realization of light DM from coannihilation, that is coupled to the  SM, predicts the presence of new light particles that can be searched for in intensity frontier experiments~\cite{Izaguirre:2014dua, Aguilar-Arevalo:2017mqx, Battaglieri:2016ggd, Izaguirre:2013uxa, Banerjee:2016tad}.

To conclude this section, it is appropriate to mention that this model has several potential sources of tuning.  There is a hierarchy problem for $\phi$, as a consequence of its $\mathcal{O}(1)$ Yukawa coupling to $\psi$  (analogous to the top quark coupling to the SM Higgs).  Therefore the $\phi$ mass has tuning of size $ \sim y^2 / (4 \pi)^2  \times m_\phi^2/\Lambda^2$, where $\Lambda$ is the high-energy cutoff of the model.  This tuning is naturally removed if the model is UV-completed to be supersymmetric or to have $\phi$ as a composite state.  We leave the exploration of such UV completions for future work.  
There is also tuning of the $\phi$ mass due to its mixing with the Higgs boson, because the $\phi$-Higgs mass matrix has an off-diagonal term of size $a_\phi v$.  This tuning has parametric size $\sim m_\phi^2/[\lambda v^2 (\tan2\theta)^2]$ and is worse than 1\% above the dashed line in Figs.~\ref{fig:phenoHiggs} and~\ref{fig:phenoHiggs2}.  This tuning is not independent of the $\phi$ hierarchy problem, since one tuning is needed to keep $\phi$ light.  We highlight this region since it is difficult to remove the tuning from singlet-Higgs mixing using a UV completion, because the tuning follows from IR parameters.  We note that the tuning from $\phi$-Higgs mixing applies generically to theories with a light scalar mixing with the Higgs~\cite{Clarke:2013aya, Krnjaic:2015mbs, Evans:2017kti}, since it does not depend on the Yukawa coupling of $\phi$ to $\psi$.

There is also a second potential tuning required to keep $\psi$ light, since $m_\psi$ gets a contribution from the VEV of $\phi$ (which is unavoidable because a tadpole is generated for $\phi$ after electroweak symmetry breaking).  This tuning has parametric size $[\lambda_\phi/(\lambda\tan2\theta)]^{1/3}m_\psi/(yv)$, and is worse than 1\% above the second dashed line in Figs.~\ref{fig:phenoHiggs} and~\ref{fig:phenoHiggs2}.

%%%%%
\begin{figure}[!t]
\begin{center}
\includegraphics[width=0.6\textwidth]{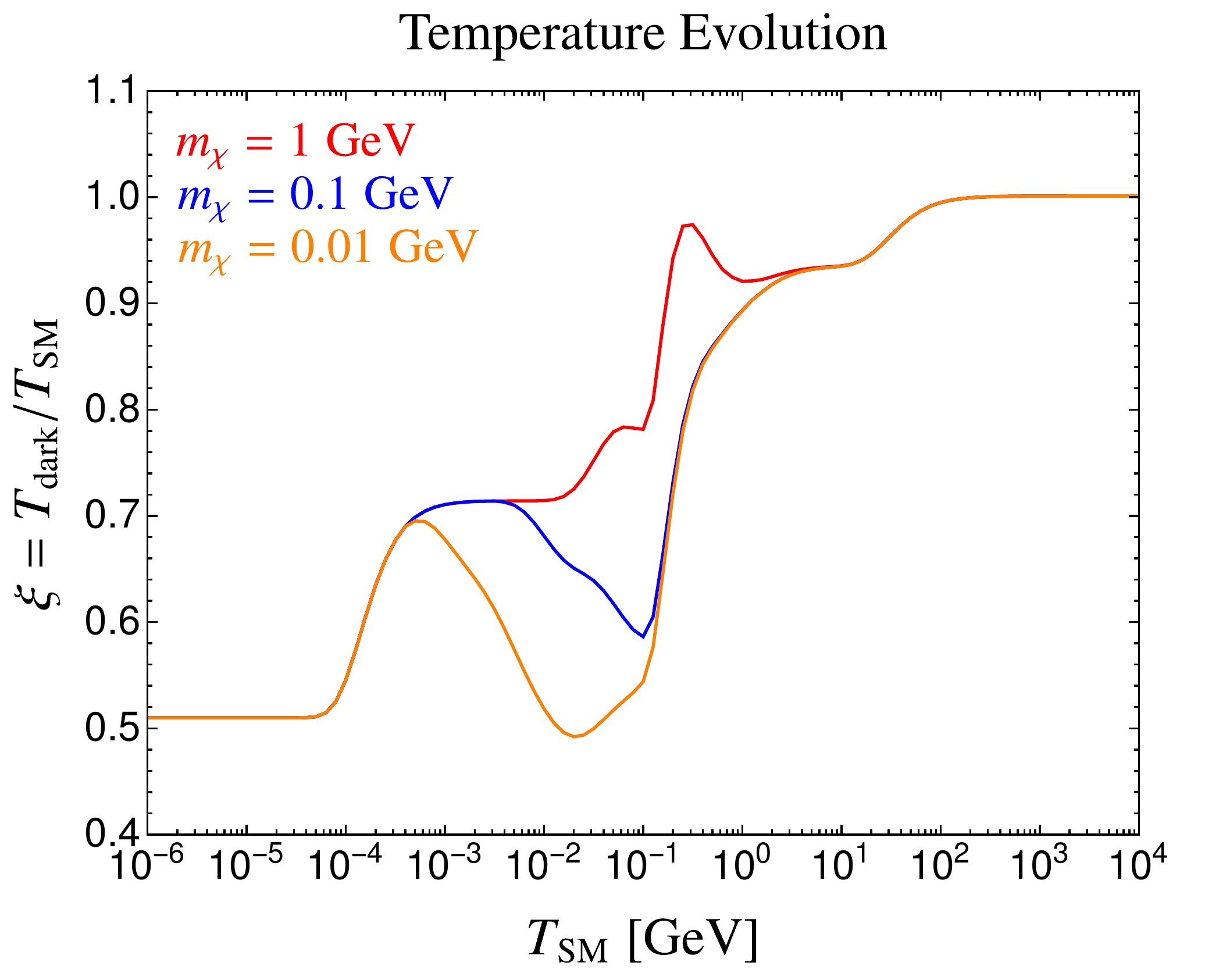}
\end{center}
\vspace{-.3cm}
\caption{Dark sector temperature evolution as a function of the SM photon temperature in the decoupled dark sector model of Sec.~\ref{sec:decoupled}.  We have assumed that the two sectors begin at the same temperature, $\xi_R = 1$, above the weak scale. For each $m_\chi$, $\Delta$ is fixed to reproduce the observed relic density. The remaining parameters are set to  $m_\phi = m_\chi/4$, $y=e^{i \pi/4}$, and  $\delta = 10^{-4}e^{i \pi/4}$.}
\label{fig:TDark}
\end{figure} 
%%%%%%

%%%%%
\begin{figure}[!t]
\begin{center}
\includegraphics[width=0.9\textwidth]{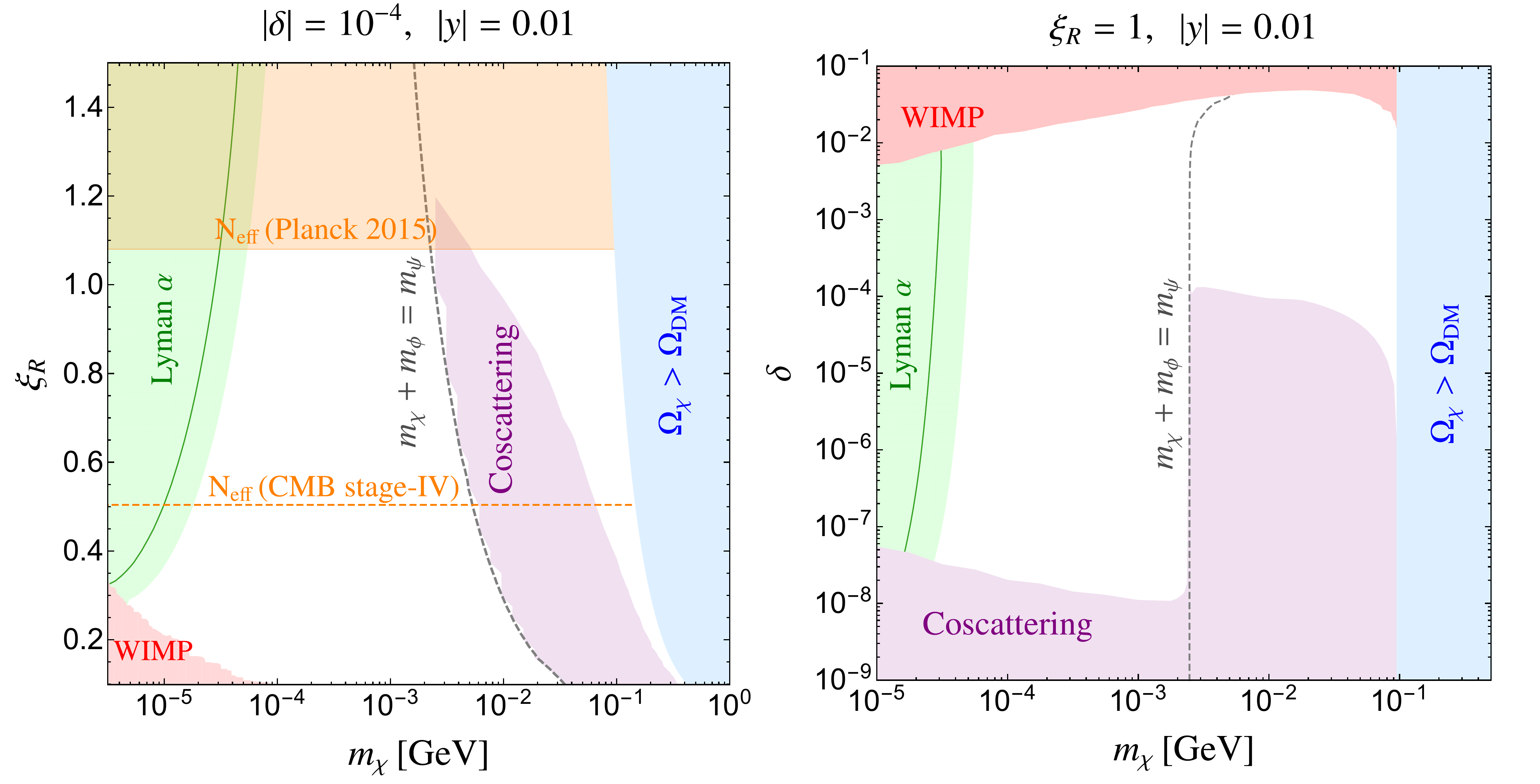}
\end{center}
\vspace{-.3cm}
\caption{Phenomenology of a dark sector decoupled from the SM, but in equilibrium with dark radiation. In the {\it left} panel we show the bound on the relative temperatures of the dark and SM sectors at reheating, $\xi_R = T_\text{dark}^R / T_{SM}^R$, as a function of the DM mass.  We show constraints on $N_{\rm eff}$ from current Planck measurements~\cite{Ade:2015xua} and the projected CMB Stage-4 sensitivity~\cite{Abazajian:2016yjj}. In the {\it right} panel we show the DM mixing angle, $\delta$,  with the active state, versus the DM mass.  When the mixing is too large or small the relic density is no longer set by coannihilation. 
The range of allowed DM masses is determined by DM overproduction and constraints on the free-streaming length from Lyman-$\alpha$ measurements~\cite{Viel:2013apy, Baur:2015jsy, Irsic:2017ixq,Yeche:2017upn}. In every point of the plot, $\Delta$ is fixed to reproduce the observed relic density. The remaining parameters are set to $y=0.01 e^{i \pi/4}$, $m_\phi = m_\chi/4$, and $\delta = 10^{-4} e^{i \pi/4}$  ($\xi_R =1$) in the {\it left} ({\it right}) panel. 
}
\label{fig:phenoDark}
\end{figure} 
%%%%%%

\subsection{Decoupled Dark Sector}\label{sec:decoupled}
%%%%%%%%%%%%%%%%%%%%%%%%%%%%%%%%%%%%%%%%%%%%%%%%%%%%%%%%%%%%%%%%%%%%%%%%%%%%%%%%%%%%%%%%%%%
%%%%%%%%%%%%%%%%%%%%%%%%%%%%%%%%%%%%%%%%%%%%%%%%%%%%%%%%%%%%%%%%%%%%%%%%%%%%%%%%%%%%%%%%%%%
Sterile coannihilation can also be realized in a completely decoupled sector.   In this case, the dark sector has its own temperature which can be colder than the SM sector, allowing the DM mass to drop below the MeV scale consistent with bounds from $N_{\rm eff}$.
 As an example we consider again the model described at the beginning of Sec.~\ref{sec:toyrelic}, supplemented with a new Majorana fermion $n$ that acts as dark radiation.  We note that our calculation of the relic density has assumed that DM is in thermal equilibrium with some form of radiation, and therefore dark radiation is required if the dark sector is decoupled from the SM\@.

A simple potential that keeps DM in thermal equilibrium with the dark radiation, $n$, is 
\be
V \supset \frac{m_n}{2}n^2+\frac{y_n}{2} \phi n^2\, .
\ee
In what follows we take $m_n \ll 1~\mathrm{eV}$, which ensures that $\Omega_n /\Omega_{DM} \ll 10\%$\@. 
We fix $y_n=10^{-4}$, which keeps $n$ in equilibrium with the dark sector.  The reheating is captured by the parameter $\xi_R = T_\text{dark}^R / T_{SM}^R$, where $T_{\text{dark},SM}^R$ are the temperatures of the two sectors at reheating.  
In what follows, we assume that the SM is reheated above the electroweak scale, $T_{SM}^R > 250~\mathrm{GeV}$.  The relative temperatures of the two sectors, $\xi = T_\text{dark} / T_{SM}$, evolves following conservation of relative entropies of the two sectors (Eq.~\ref{eq:entropy}), and depends on the number of relativistic degrees of freedom in each sector.  Fig.~\ref{fig:TDark} shows the temperature evolution, assuming $\xi_R=1$, for a few representative dark spectra.  We see that the dark sector ends up cooler than the SM, when it starts with the same temperature, as a consequence of the larger number of degrees of freedom in the SM sector. 

Even if only gravitationally coupled to the SM, there are several ways that the dark sector can be probed observationally.
 First, $N_{\rm eff}$ during BBN or the CMB epoch can be modified. Second, if the DM mass is sufficiently small, the matter power spectrum can deviate from the $\Lambda$CDM prediction~\cite{Ade:2015xua, Spergel:2003cb, Hannestad:2003ye, Tegmark:2003ud, Crotty:2004gm, Seljak:2004xh, Bond:1980ha, Lesgourgues:2006nd, Viel:2005qj, Hooper:2007tu}. Third, DM self-interactions can leave imprints on astrophysical observations~\cite{Spergel:1999mh, Markevitch:2003at, Clowe:2003tk, Randall:2007ph,Rocha:2012jg, Peter:2012jh, Harvey:2015hha}. However, in sterile coannihilation, self-interactions are generically suppressed by the small couplings ($\delta$ in our case) of DM to the mediator.  In our model, the DM self-interaction cross section is smaller than weak scale, whereas observable self-interactions requires a nuclear sized cross section.

In the left panel of Fig.~\ref{fig:phenoDark} we show the current bound from $N_{\rm eff}$, as measured by Planck~\cite{Ade:2015xua}, on the reheating temperature of the dark sector. A dark sector reheated slightly above the SM is still consistent with observations. We also display the projected sensitivity of CMB Stage-4~\cite{Abazajian:2016yjj}. BBN bounds on $N_{\rm eff}$ are subleading. 

In the decoupled dark sector, the smallest DM mass consistent with observations is determined by Lyman-$\alpha$ measurements~\cite{Viel:2013apy, Baur:2015jsy, Irsic:2017ixq,Yeche:2017upn}. Two effects can suppress the matter power spectrum: the free-streaming of DM~\cite{Lesgourgues:2006nd,Hooper:2007tu} and dark acoustic oscillations~\cite{Loeb:2005pm,Hooper:2007tu,Bringmann:2016ilk, Binder:2016pnr}. We find that the second effect is subdominant in our parameter space, and the lower bound on the DM mass is set by the free-streaming length. 

After decoupling from dark radiation, DM particles stream freely due to their non-zero velocity dispersion, damping density perturbations below a certain scale.  This free-streaming scale is given by the comoving length scale that a DM particle can travel between the time of kinetic decoupling and matter radiation equality.
DM is kept in kinetic equilibrium with dark radiation through (inverse-)decay, $\psi \leftrightarrow \chi \phi$, where $\phi$ is in thermal contact with dark radiation.  For parameters where the free streaming bound is relevant, we find that decays decouple later than scatterings (such as $\chi \phi \rightarrow \chi \phi$).
We estimate the temperature of the SM bath at DM kinetic decoupling, $T_{SM}^\text{kd}$, using the relation $m_\psi\Gamma_\psi \left[n_\psi/(\rho_\chi+\rho_\psi)\right]_{T_{SM}^\text{kd}} = H(T_{SM}^\text{kd})$. 
The free-streaming length is 
\bea
\lambda_{\text{FS}} \quad =\quad  \int_{t_\text{kd}}^{t_\text{EQ}}\frac{v(t)}{a(t)}dt\nn & \approx & \frac{1}{a(T_\text{EQ})}v_\text{kd}\frac{3\sqrt{5}M_\text{Pl}}{2\pi^{3/2}g_*(T_\text{EQ})^{1/2}T_\text{EQ}T_{SM}^\text{kd}}\log\left(\frac{T_{SM}^\text{kd}}{T_\text{EQ}}\right)
\nn \\ 
&\approx& 0.124 \text{ Mpc}\ v_\text{kd} \frac{\text{keV}}{T_{SM}^\text{kd}}\log\left(\frac{1.4\ T_{SM}^\text{kd}}{\text{eV}}\right) \, ,
\eea
where $v_\text{kd}\sim\sqrt{T_\text{dark}^\text{kd}/m_\chi}$ is the velocity of DM particles at kinetic decoupling, $T_\text{EQ}$ is the SM temperature at matter radiation equality, and we have approximated the number of relativistic degrees of freedom, $g_*$, as constant between $T_{SM}^\text{kd}$ and $T_\text{EQ}$.

 In Fig.~\ref{fig:phenoDark}, we show the bound for $\lambda_{\text{FS}} \lesssim  0.06$ Mpc~\cite{Irsic:2017ixq} (green shaded area) and $\lambda_{\text{FS}} \lesssim $ 0.1 Mpc~\cite{Viel:2013apy} (green line). When the dark sector is reheated to the same temperature as the SM, DM masses below $\sim 100$ keV are excluded. The bound becomes weaker if the dark sector is colder than the SM, as shown in the left panel of Fig.~\ref{fig:phenoDark}, and we find viable sterile coannihilation models with the DM as light as $m_\chi \approx 5$~keV\@.  We find that the constraint from Lyman-$\alpha$ is stronger than the Tremaine-Gunn bound~\cite{Tremaine:1979we}, which sets a lower limit on the mass of fermionic DM of several hundred eV~\cite{Boyarsky:2008ju}.

The rest of our parameter space is bounded by regions where coannihilation is not setting the relic density. At large $\delta$ or small $\xi_R$, $\chi\chi$ annihilations dominate freeze-out and DM becomes WIMP-like. When $\delta$ is too small, processes that exchange $\chi$ and $\psi$ decouple before annihilations, and freeze-out enters the coscattering phase~\cite{DAgnolo:2017dbv}. Finally, at large $m_\chi$ (and fixed $y$), the relic density is too large because the effective annihilation cross section, which scales as $y^4/m_{\chi}^2$, becomes smaller than weak scale for any choice of $\Delta$.

%%%%%%%%%%%%%%%%%%%%%%%%%%%%%%%%%%%%%%%%%%%%%%%%%%%%%%%%%%%%%%%%%%%%%%%%%%%%%%%%%%%%%%%%%%%
%%%%%%%%%%%%%%%%%%%%%%%%%%%%%%%%%%%%%%%%%%%%%%%%%%%%%%%%%%%%%%%%%%%%%%%%%%%%%%%%%%%%%%%%%%%
\section{Direct Annihilations to Standard Model Particles} \label{sec:SM}
Coannihilations consisting of direct annihilations to SM particles offer unique detection prospects in future experiments.  Existing studies have highlighted thermal targets, where masses and couplings are chosen such that the relic density matches the observed abundance of DM\@.  Several studies have considered coannihilation, but focus on models where DM participates in the dominant annihilation channel~\cite{Izaguirre:2017bqb, Feng:2017drg, Battaglieri:2017aum, Darme:2017glc}.  In these models, the lifetime of heavy states to decay to DM is related to the DM relic density, providing predictive targets for future experiments that probe this lifetime.

In this section, we highlight the sterile coannihilation regime, where the heavy state annihilates directly to the SM\@.  The relic density is independent of the DM couplings, since DM does not participate in the dominant annihilations.  Therefore, sterile coannihilation points to a broad parameter space of thermal targets, where the lifetime of the heavy states can be varied without changing the relic density.  If the lightest state is DM today, the only robust requirement is that the lifetime of the heavy states should be shorter than the present age of the Universe.

%%%%%
\begin{figure}[!t]
\begin{center}
\includegraphics[width=0.9\textwidth]{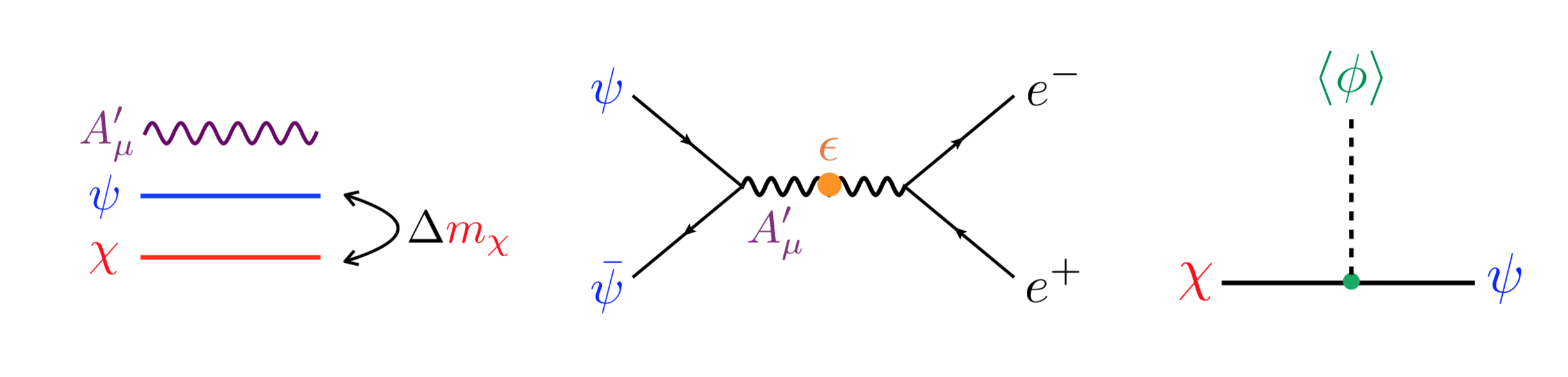}
\end{center}
\vspace{-.3cm}
\caption{Schematic representation of the model described in Sec.~\ref{sec:SM}. From left to right: mass spectrum, main annihilation channel, and mixing between DM, $\chi$, and its coannihilating partner, $\psi$.}
\label{fig:SMLayout}
\end{figure} 
%%%%%%

We  illustrate this point using the following model: the dark sector contains two Weyl spinors, $\psi$ and $\psi^c$, charged under a dark gauge group, $U(1)_{D}$, with gauge boson $A'_{\mu}$ and gauge coupling $g_D\equiv \sqrt{4\pi\alpha_D}$. The gauge symmetry is spontaneously broken by a complex scalar $\phi$ with VEV $\left< \phi \right> \equiv v_\phi$. Gauge charges are $+1$ for $\psi$ and $-1$ for $\phi$ and $\psi^c$. We include an additional neutral Majorana spinor, $\chi$, which is our DM candidate. The fermionic part of the dark sector potential is 
\bea
V &\supset & m_\psi \psi^c \psi + \frac{m_\chi}{2}\chi^2  + y\phi \psi \chi + \bar y \phi^\dagger \psi^c \chi+ {\rm h.c.}\, ,
\eea
where all parameters ($m_{\psi}$, $m_{\chi}$, $y$, and $\bar{y}$) are complex. We can remove 3 phases, by redefining the fields, so we are left with 4 real parameters and 1 phase. We consider the  sterile coannihilation limit, $y v_\phi/m_{\psi, \chi}\ll 1$ and $\bar y v_\phi/m_{\psi, \chi}\ll 1$, such that $\chi$ corresponds to DM, while $\psi$ and $\psi^c$ correspond to the heavy states whose annihilations dominate the effective cross section of Eq.~\ref{eq:relicco}. In this example, we take $m_\psi \geq m_\chi$ and define $\Delta \equiv (m_\psi - m_\chi)/m_\chi$ as the approximate mass splitting between DM and its two coannihilating partners (that are almost degenerate).  We assume that $m_\phi > m_{\chi, \psi}$, so that annihilations into $\phi$ can be neglected.

We consider the regime $m_{A^\prime} > m_{\psi, \chi}$,  and we assume that there is a kinetic mixing between the new gauge group $U(1)_D$ and electromagnetism
\be
\mathcal{L} \supset - \frac{\epsilon}{2}F_d^{\mu\nu} F_{\mu\nu}\, ,
\ee
where $F_d$ is the field strength of the dark gauge force and $\epsilon$ sets the strength of the mixing. 

The spectrum contains three fermionic mass eigenstates $n_{1,2,3}$, which we order by ascending mass ($m_{n_i} < m_{n_j}$ if $i<j$).   $n_1$ is mostly $\chi$ and is weakly interacting and $n_{2,3}$ are mostly made up of $\psi$ and $\psi^{c\dagger}$. The main ingredients of the model are summarized in Fig.~\ref{fig:SMLayout}. As shown in the figure, coannihilation proceeds through an off-shell dark photon into SM particles. Kinetic equilibrium with the SM is insured by dark photon decays and DM scattering off SM states. The only $\mathcal{O}(g_D)$ coupling after mass diagonalization pairs up $n_2$ and $n_3$ in the dark photon vertex. The other couplings are suppressed by powers of $\delta \equiv y v_\phi/m_{\chi}$ and $\bar \delta \equiv \bar y v_\phi/m_{\chi}$. Therefore, as for the toy model in Sec.~\ref{sec:toyrelic}, so long as $\delta$ and $\bar \delta$ are sufficiently small, the relic density is independent of them, since 
\be
\langle \sigma_{\rm eff} v \rangle = \langle \sigma_{23} v \rangle +\mathcal{O}(\delta^2, \bar \delta^2)\, ,
\ee
where $\sigma_{23}$ is the annihilation cross section of $n_2 n_3$ to SM states. 

The phenomenology of the model  strongly depends on  $\delta$ and $\bar \delta$, for two reasons. First, $n_1$ interactions are suppressed by $\delta$ and $\bar \delta$, and in most of the parameter space DM today is entirely composed of this lightest state. Second, the heavier and more strongly interacting fermions, $n_{2,3}$, have decay widths proportional to these small parameters, $\Gamma_{n_{2,3}}\sim \delta^2, \bar \delta^2$.

\begin{figure}[!t]
\begin{center}
\includegraphics[width=0.98\textwidth]{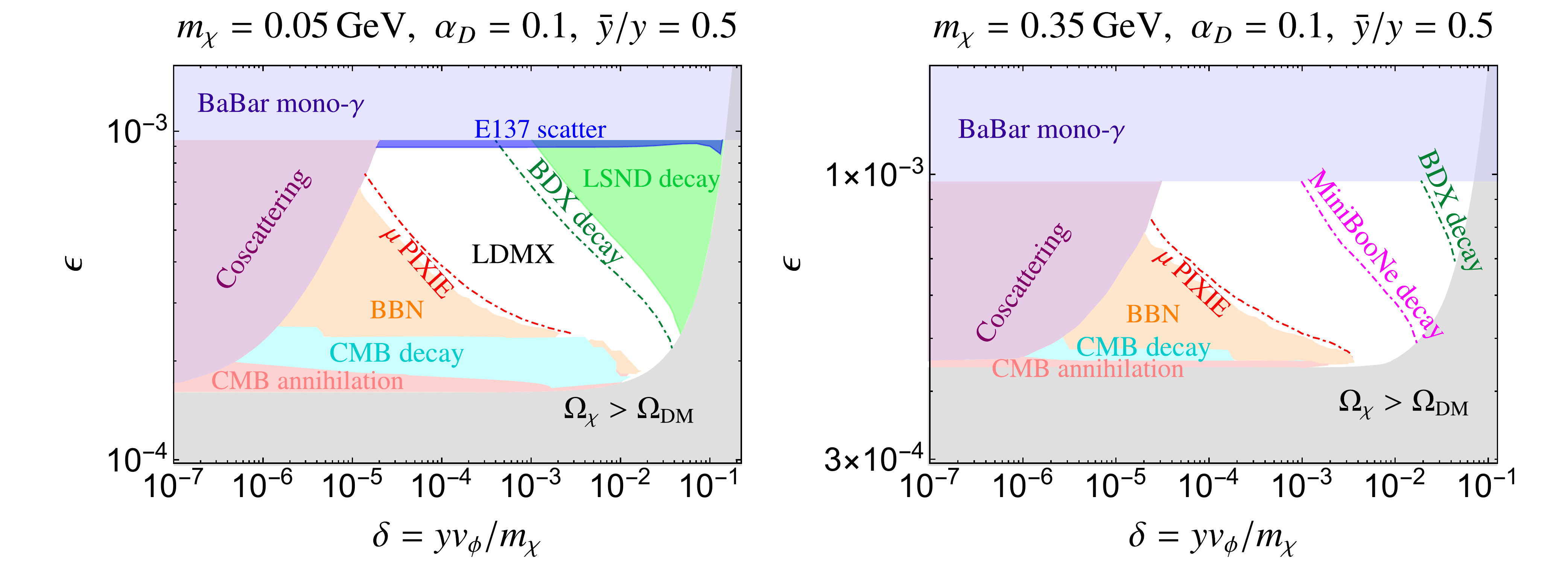}
\end{center}
\vspace{-.3cm}
\caption{Dark photon coupling, $\epsilon$, vs. the size of the DM mixing with its coannihilating states, $\delta$. When $\delta$ is sufficiently small the relic density does not depend on it, as shown by the flat $\Omega_\chi>\Omega_{\text{DM}}$ contour. On the contrary the lifetime of the heavier states, sharing a conserved quantum number with DM, increases as $\delta^2$, strongly affecting the phenomenology. 
The bounds are discussed in the text and come from BaBar~\cite{Essig:2013vha, Lees:2017lec}, E137~\cite{Bjorken:1988as}, LSND~\cite{ Auerbach:2001wg}, BBN~\cite{Jedamzik:2006xz, Poulin:2015opa}, and the CMB~\cite{Poulin:2016anj}. The reach for BDX and MiniBooNe adapted from~\cite{Izaguirre:2017bqb}, and PIXIE~\cite{Poulin:2016anj, Chluba:2013pya, Kogut:2011xw, Abitbol:2017vwa} (corresponding to $\mu < 5 \times 10^{-8}$) are shown. LDMX~\cite{Izaguirre:2014bca} can probe the entire allowed region of the left panel, but is not sensitive to the heavier dark photon mass on the right panel. In every point of the plot $\Delta$ is fixed to reproduce the observed relic density. The remaining parameters are set to $m_{\chi}=0.05\ (0.35)$ GeV, $m_{A'}=3 m_{\chi}\left(1+\Delta \right)$, $\alpha_{D}=0.1$, and $\bar{y}/ y = 0.5$ in the {\it left}  ({\it right})  panel.}
\label{fig:SMann}
\end{figure}

In the laboratory, $n_{2,3}$ are the the states that are dominantly produced, and their decay length, into $n_1$, determines whether they can be detected or not. A quantitative illustration of this point is presented in Fig.~\ref{fig:SMann}. There we fix the dark sector mass scale, $m_\chi$, and dark gauge coupling $\alpha_D$, and look at the reach on the dark photon coupling to the SM, $\epsilon$, as a function of the parameter $\delta$.  For simplicity we take $\bar \delta = \delta/2$. At every point $\Delta$ is chosen to reproduce the observed relic density.\footnote{For the DM mass chosen in the right panel of Fig.~\ref{fig:SMann}, DM annihilations into hadronic final states play an important role and are not treated by \texttt{MicrOMEGASv4}, which works in the partonic limit.  In this case, we solve the Boltzmann equation numerically,  and we take into account annihilation into hadrons by replacing the annihilation cross section into leptonic final states with $(\sigma v)_{e^+e^-}+(\sigma v)_{\mu^+\mu^-}\left[1+R(s)\right]$, where $(\sigma v)_{e^+e^-}$ and $(\sigma v)_{\mu^+\mu^-}$ are the annihilation cross sections of DM into $e^+e^-$ and $\mu^+\mu^-$ respectively and $R(s)$ is the $R$-ratio for $e^+e^-$ hadronic annihilation. We use the $R(s)$ values provided by the PDG~\cite{Beringer:1900zz}.}

%%%%%
%%%%%%

We show in Fig.~\ref{fig:SMann} the bound from the LSND~\cite{Auerbach:2001wg} proton beam dump and the projected reach of the BDX~\cite{Battaglieri:2014qoa} and MiniBooNe~\cite{Bazarko:2000id, AguilarArevalo:2008qa} electron beam dumps.  The sensitivity of these experiments comes from the production of the dark photon, $A^\prime$, which subsequently decays promptly with a nearly $100\%$ branching ratio to $n_2 n_3$. Then if at least one of the two states decays within the active volume of the detector, the leptons from $n_{2, 3}\to e^+ e^- n_1$ can be detected.   The reach in $\epsilon$ of these three beam dumps depends on the lifetime of $n_2$ and $n_3$, and therefore on $\delta$.  To derive the bounds and reach, we used the results of Ref.~\cite{Izaguirre:2017bqb}, after accounting for the different lifetime of the excited state in our model.  We include a factor of 2 enhancement to our signal rate, compared to Ref.~\cite{Izaguirre:2017bqb}, because each dark photon decay produces two excited states.  

We find that the proposed LDMX electron beam dump~\cite{Izaguirre:2014bca} can probe the entire allowed region of the  left panel of Fig.~\ref{fig:SMann}.   In our model the LDMX signal will be missing energy events, from the production of an $A^\prime$, followed by its decay to $n_2 n_3$ (surviving past the detector calorimeters).  In the right panel of Fig.~\ref{fig:SMann}, the dark photon is too heavy to be produced at LDMX so there is no sensitivity.

Even after fixing the relic density, the lifetime of $n_2$ and $n_3$ can be large enough that decays during BBN~\cite{Jedamzik:2006xz, Poulin:2015opa} and during recombination~\cite{Poulin:2016anj} become relevant.  Decays of the heavier states can also leave CMB spectral distortions detectable by the proposed PIXIE satellite~\cite{Poulin:2016anj, Chluba:2013pya, Kogut:2011xw, Abitbol:2017vwa}. 
Spectral distortions are a genetic signal of sterile coannihilation, because the heavier coannihilating states have a long lifetime in a large fraction of the parameter space.    This is due to the insensitivity of the relic density to the small mixing between DM and the active states.  
We note that a signal in LDMX, plus the detection of  spectral distortions, would point to DM from sterile coannihilation.

In the region where $\Delta$ approaches zero, $n_2$ and $n_3$ can live long enough that they are still present at the time of recombination. When this happens, their annihilations can alter the CMB temperature and polarization power spectra.  We show the corresponding bound from Planck~\cite{Ade:2015xua} in Fig.~\ref{fig:SMann}.

The remaining constraints that we show in Fig.~\ref{fig:SMann} are a monophoton search from BaBar~\cite{Essig:2013vha, Lees:2017lec}, and the electron beam dump E137~\cite{Bjorken:1988as}. In the latter experiment, $n_2$ and $n_3$ are produced through an on-shell dark photon, as in the previous cases, and detected through scattering off electrons $n_{2,3} e^- \to n_{3,2} e^-$.

We have seen that sterile coannihilation, with annihilations of the heavy state directly into the SM, has a rich phenomenology.  Sterile coannihilation is generically described by a weakly interacting DM candidate accompanied by heavier, more strongly interacting, states that can have cosmologically long lifetimes.  Sterile coannihilation leads to a broader parameter space of lifetimes, consistent with the observed DM abundance, than the thermal targets of non-sterile coannihilating models.

\section{Conclusions} \label{sec:conclusions}
%%%%%%%%%%%%%%%%%%%%%%%%%%%%%%%%%%%%%%%%%%%%%%%%%%%%%%%%%%%%%%%%%%%%%%%%%%%%%%%%%%%%%%%%%%%
%%%%%%%%%%%%%%%%%%%%%%%%%%%%%%%%%%%%%%%%%%%%%%%%%%%%%%%%%%%%%%%%%%%%%%%%%%%%%%%%%%%%%%%%%%%
In this paper, we discussed how coannihilation constitutes a novel mechanism for light DM that evades CMB constraints on WIMPs.
Traditionally, coannihilation has been applied to weak scale DM, where it is only relevant for highly degenerate states.    As we have emphasized, non-degenerate coannihilation naturally realizes light DM\@. This is because the annihilation rate of heavy states is exponentially suppressed, by their equilibrium number density, when computing the effective annihilation rate that dictates the DM relic density (see Eq.~\ref{eq:relicco}).  If the annihilations of heavy states dominate, they must have a cross section that is exponentially larger than the weak scale, in order to overcome this exponential suppression.  To generate a large cross section, the mass scale of the dark states should be exponentially lighter than the weak scale. The CMB bound is evaded simply because the heavy states can decay to DM before recombination, such that the annihilations that dictate the relic density stop occurring before the CMB forms.

We have focused on the sterile coannihilation limit, where DM does not participate in the dominant annihilations.  The relic density depends on the DM mass, but not its couplings.  This is a counterexample to the lore that the DM's couplings are determined by its abundance, once its mass is specified.  Therefore, sterile coannihilation leads to a broader parameter space of thermal targets for future experiments than is usually considered.

Light DM from coannihilation is easily realized in a variety of dark sectors.  We have studied three example models, where mostly sterile DM mixes with a heavier state that experiences rapid annihilations.  In Sec.~\ref{sec:decoupled}, we considered the case that the dark sector contains dark radiation and is totally decoupled from the SM sector.  We found that the DM can be as light as the keV scale.  In Sec.~\ref{sec:secluded}, we considered DM that is coupled to the SM and annihilates into a dark Higgs that mixes with the SM Higgs.    This possibility is more minimal than the decoupled scenario, because it does not require dark radiation and does not depend on the initial DM temperature.  The lower bound on the DM mass is raised by three orders of magnitude, compared to the decoupled case, due to constrains on $N_{\rm eff}$ from BBN and the CMB\@.  We identified rich experimental prospects for production and detection of the dark states.  In Sec.~\ref{sec:SM}, we considered direct annihilations to the SM in the limit of sterile coannihilation.  This scenario provides an explicit example of thermal targets that are independent of DM's couplings and therefore span a wide range of lifetimes for the heavy states to decay\@.  Sterile coannihilation points to a characteristic combination of signals in future missing momentum experiments and CMB spectral distortions. 

We note that light  DM from coannihilation is a broad framework that extends beyond the specific example models considered in this paper.  It would be interesting to look at more example dark sectors exhibiting light DM from coannihilation, and more portals that connect them to the SM\@.  It is worthwhile to develop explicit models where the dark sector is supersymmetric, or where the DM and annihilating states are composites of a strongly coupled hidden sector, such that there masses are naturally understood. Finally, we note that coannihilation can be combined with other mechanisms, such as forbidden annihilations, dark sector cannibalism, and 3-to-2 annihilations.

\section*{Acknowledgements}

The authors would like to thank  Francesco D'Eramo, Daniel Egana-Ugrinovic, Yonit Hochberg, Duccio Pappadopulo, Gabriele Trevisan, Neal Weiner, and Mike Williams for helpful discussions. RTD is supported by the U.S. Department of Energy under Contract No.~DE-AC02-76SF00515. JTR is supported by NSF CAREER grant PHY-1554858. CM would like to thank the CERN Theoretical Physics Department for hospitality while this work was completed.  RTD and JTR thank the hospitality of the Aspen Center for Physics, which is supported by the NSF grant PHY-1607611.

\end{document}